\documentclass[useAMS,usenatbib,usegraphicx,floatfix,epsfig.epsf]{mn2e}
\usepackage[letterpaper,hmargin=1in,vmargin=1.25in]{geometry}
\usepackage{graphicx}
\usepackage{amsmath}
\usepackage{bm}
\usepackage{amssymb}

{\catcode`\|=\active 
  \gdef\Braket#1{\begingroup
     \ifx\SavedDoubleVert\relax
       \let\SavedDoubleVert\|\let\|\BraDoubleVert
     \fi
     \mathcode`\|32768\let|\BraVert
     \left<{#1}\right>\endgroup}
}
\begin{document}

\title[Trans-Neptunian binary production]
{Production of trans-Neptunian binaries through chaos-assisted capture}
\author[Lee, Astakhov, Farrelly]{Ernestine A. Lee,$^{1}\thinspace$\thanks{E-mail: ernestine.lee@gmail.com}
Sergey A. Astakhov,$^{2,3}\thinspace$\thanks{E-mail: astakhov@gmail.com}
and David Farrelly$^{2}\thinspace$\thanks{E-mail: 
david.farrelly@gmail.com}
\\
$^{1}$FivePrime Therapeutics, 1650 Owens Street, Suite 200, San
Francisco, CA 94158-2216, USA\\
$^{2}$Department of Chemistry and Biochemistry, Utah State 
University, Logan, UT 84322-0300, USA\\
$^{3}${UniqueICs, Stroiteley 1, Saratov, 410044, Russia\thinspace \thanks{present address}}\\
}

\date{}
\maketitle

\begin{abstract}
The recent discovery of binary objects in the Kuiper-belt opens an invaluable window into past and present conditions in the trans-Neptunian part of the Solar System. For example, knowledge of how these objects formed can be used to impose constraints on planetary formation theories. We have recently proposed a binary-object formation model based on the notion of chaos-assisted capture. In this model two potential binary partners may become trapped for long times inside chaotic layers within their mutual Hill sphere. The binary is then captured permanently through gravitational scattering with a third ``intruder'' body. The creation of binaries having similarly sized partners is an {\it ab initio} prediction of the model which also predicts large binary semi-major axes and moderately eccentric mutual orbits similar to those observed. Here we present a more detailed analysis with calculations performed in the spatial (three-dimensional) three- and four-body Hill approximations. It is assumed that the potential binary partners are initially following heliocentric Keplerian orbits and that their    
relative motion becomes perturbed as these objects undergo close 
encounters. First, the mass, velocity, and 
orbital element distributions which favour binary formation are identified in the circular and elliptical Hill limits. We then consider intruder scattering in the circular Hill four-body problem and find that the chaos-assisted capture mechanism is consistent with observed, apparently randomly distributed, binary mutual orbit inclinations. It also predicts asymmetric distributions of retrograde {\it versus} prograde orbits. The time-delay induced by chaos on particle transport through the Hill sphere is analogous to the formation of a resonance in a chemical reaction. Implications for binary formation rates are considered and the ``fine-tuning'' problem recently identified by \citet{Noll:2007} is also addressed.
\end{abstract}


\section{Introduction}
\label{intro}

The discovery of the first trans-Neptunian - or Kuiper-belt\footnote{The existence of a populated trans-Neptunian part of the Solar System, variously called the Kuiper-belt or the Edgeworth-Kuiper-belt, was apparently first speculated on by \citet{Leonard:1930}. After later conjectures by \citet{Edgeworth:1943,Edgeworth:1949} and \citet{Kuiper:1951},  \citet{Fernandez:1980} provided the first quantitative discussion of the possibility of a comet-belt beyond Neptune using a Monte Carlo method - see \citet{Jewitt:1999,Luu:2002,Jewitt:2006}.} - binary 
came in two parts punctuated by a prolonged hiatus; the erstwhile 
planet Pluto was discovered by Clyde Tombaugh in 1930 \citep{Slipher:1930,Basri:2006,Soter:2006} whereas its companion, 
Charon, remained undetected until 1978 when it was discovered almost by chance \citep{Christy:1978}. The 
identification of the second binary trans-Neptunian object (TNO) was also a two-step - and similarly fortuitous - process; during the analysis of orbit recovery data for 1998 WW$_{31}$ it was recognized that this object is actually a binary \citep{Veillet:2002}. 
This discovery sparked observing campaigns designed to discover other trans-Neptunian binaries (TNBs) \citep{Brown:2002} and, since then, the inventory 
of TNBs has expanded rapidly; including Centaurs \citep{Noll:2006b}, currently about 30 such binaries are known \citep{Johnston:2007,Noll:2003,Osip:2003,Kern:2006,Kern:2006a,Noll:2006,Noll:2006a,Noll:2006b,Noll:2007,Stansberry:2006,Stephens:2006}

In planetary physics, as in other areas of astronomy \citep{Noll:2003}, the motivation for finding binaries is that they can be recruited as natural probes of past and present conditions in their locale - in this instance the Kuiper-belt \citep{Stern:1996,Williams:1997,Kenyon:1998,Jewitt:1999,Farinella:2000,Malhotra:2000,Allen:2001,Kenyon:2002,Luu:2002,Schulz:2002,Levison:2003,Jones:2005,Chiang:2006,Levison:2006,Morbidelli:2006}. For example, knowledge of binary mutual orbits provides for a direct and accurate measurement of the total mass of the system. This then paves the way for the determination of other properties, e.g., bulk densities and mass distributions \citep{Toth:1999,Noll:2003,Hestroffer:2005,Noll:2006a,Kenyon:2004,Cruikshank:2006,Descamps:2007}. These data, together with information about binary frequencies, are key to understanding the formation and evolution of accretion disks \citep{Luu:2002,Cruikshank:2006}.

Amongst Solar System binaries \citep{Merline:2002,Noll:2003}, TNBs are of particular interest because of their rather unusual orbital and physical properties. These include large, moderately eccentric, mutual orbits; randomly distributed inclinations; and a seeming preference for the binary partners to have comparable sizes \citep{Margot:2002,Burns:2004,Noll:2003,Jones:2005,Noll:2006a,Noll:2007}.  The discovery of peculiar properties is a felicitous event because these observations can be used to tighten constraints on theories of the formation and evolution of the Kuiper-belt and, more generally, the Solar System. For example, the distribution of binary mutual orbit inclinations potentially provides insight into the velocity dispersion of bodies in the primordial Kuiper-belt \citep{Chiang:2006,Noll:2007,Goldreich:2004}.

The implication of the discovery of large binary orbits with roughly equal mass partners is that TNBs did not form through physical collisions; generally these are expected to produce objects with rather asymmetric masses and relatively small orbits \citep{Margot:2002,Burns:2004,Durda:2004,Stern:2002}. However, an important caveat applies; TNO binaries are difficult to observe even with the largest ground-based telescopes \citep{Toth:1999} which opens up the possibility that the apparent preference for large, symmetric-mass binaries is, in reality, the result of observational bias \citep{Burns:2004}. Fortunately,  it has been possible significantly to better characterize these objects with the Hubble Space Telescope (HST) High Resolution Camera (HRC): \citet{Noll:2006a} have recently made very deep observations of TNBs using the HRC in which they determined magnitude differences, $\Delta_{mag}$, between binary components.  These are the first observations capable of measuring the relative frequency of symmetric and asymmetric binaries, albeit with the assumption that relative brightness is a proxy for size.  \citet{Noll:2006a} observed statistically significant clustering of binaries with  $\Delta_{mag} < 1$ and concluded that the preference for symmetric binaries is probably real and peculiar to TNBs. 

We have recently proposed a dynamical TNB formation mechanism \citep{Astakhov:2005} based on the idea of chaos-assisted capture (CAC) in the circular Hill problem \citep{Astakhov:2003,Astakhov:2004}. Chaos-assisted capture happens because the interface between bound and scattering regions of phase space in the Hill problem consists of ``sticky'' chaotic layers in which trajectories can become trapped and mimic regular (i.e., non-chaotic) orbits for long periods of time \citep{Perry:1994,Simo:2000,Zaslavsky:1985,Astakhov:2003}. The extension of the lifetime of the transient binary through its entanglement in chaotic layers then provides the opportunity for permanent capture. We proposed that the binary is stabilized through gravitational scattering with a smaller ``intruder'' particle. Subsequent intruder scattering events gradually reduce the size of the binary orbit and this process eventually results in an essentially Keplerian mutual orbit. 

Numerical simulations in the CAC model \citep{Astakhov:2005} indicated that symmetric binaries (i.e., binaries consisting of similar sized partners) appear to be created preferentially. This was explained as being the result of chaos preferentially destabilizing asymmetric mass binaries as compared to symmetric binaries during encounters with intruders. 

Alternative TNB formation models include: physical collisions of two objects which then fuse into a single object; because all of this is assumed to take place inside the Hill sphere\footnote{a region wherein mutual gravity dominates solar differential gravity - see Table 1} of a third object a binary eventually results \citep{Weidenschilling:2002}; dynamical friction \citep{Goldreich:2002}; gravitational scattering \citep{Goldreich:2002}; and exchange reactions \citep{Funato:2004}.  Discussion of these models can be found in, e.g., \citet{Noll:2006,Noll:2007,Astakhov:2005,Kern:2006,Cruikshank:2006}. Generally, physical collisions alone are unlikely to have formed binaries with large mutual semi-major axes, in part, because TNBs have significantly more angular momentum than typically results from a collision \citep{Margot:2002,Burns:2004,Durda:2004,Stern:2002,Canup:2005,Chiang:2006} although the satellites of Pluto and of some other TNOs likely have a collisional origin \citep{Canup:2005,Brown:2007,Morbidelli:2007}.  This suggests that perhaps the majority of TNBs have a dynamical origin, e.g., involving gravitational scattering or dynamical friction \citep{Kern:2006}. The CAC model invokes scattering inside the Hill sphere of three small bodies - the potential binary partners (the ``primaries'') and a third, intruder particle. Thus, the overall process is four-body including the Sun. 

Here we present a more detailed investigation of the initial stages of capture in the CAC scenario, initially in the spatial three-body circular and elliptical Hill approximations. In part this is an attempt to understand how the asymptotic pre-encounter orbital elements determine capture probabilities. Subsequently, we focus on how the mechanism in the four-body circular Hill problem, in particular, depends on the masses and velocities of the three small bodies. 

A note on terminology is in order: prior to permanent capture a temporary binary must form. This object will be referred to as a transient, quasi-bound, or proto-, binary. Immediately after capture the binary typically has a very large non-Keplerian orbit; henceforth we refer to it as a ``nascent'' binary. The mechanics of orbit reduction of the nascent binary by further intruder scattering events is not considered in detail in this article and will be reported separately. For convenience the main symbols and terms used and their definitions are collected together in Table 1.

The paper is organized as follows: Section \ref{tbhill} introduces the Hamiltonian and equations of motion of the elliptic spatial Hill problem. Also in Sec.\ \ref{tbhill} we briefly review the CAC mechanism and define orbital elements suitable for describing the motion of the primaries at infinity in the Hill approximation. A similar approach has been employed in studies of collisions between planetesimals in a disk revolving around the Sun \citep{Wetherill:1989,Greenzweig:1990,Greenzweig:1992,Wetherill:1993,Nakazawa:1989,Ohtsuki:1990,Ohtsuki:2002,Stewart:2000} or dust grains in planetary rings \citep{Petit:1987,Petit:1987a}. Distributions of orbital elements at infinity which can lead to capture in the spatial three-body circular and elliptical Hill approximations are then computed. Four-body intruder scattering simulations in the circular Hill approximation are described in Sec.\ \ref{stab4}. Results are presented in Sec.\ \ref{results}. Comparison is made with the predictions and assumptions of the models of \citet{Goldreich:2002} in Sec.\ \ref{discussion}. The binary hardening mechanism is briefly considered in Sec.\ \ref{hard}; limitations of our calculations are considered in Sec.\ \ref{limit} and conclusions are in Sec.\ \ref{conclusions}.

\begin{table*}
\begin{center}
\caption{Main symbols and terms used and their definitions}
\begin{tabular}{lll}
\hline\hline
&  &  \\ 
\textbf{Symbol/Term} &  & \textbf{Definition} \\ 
&  &  \\ \hline\hline
&  &  \\ 
$G$ &  & Gravitational constant \\ 
$M_{\odot }$ &  & Solar mass \\ 
$m_{1},m_{2}$ &  & Binary partner masses - the
``primaries'' \\ 
$m_{3}$ &  & Mass of fourth body - the intruder - scattered by primaries  \\ 
$D_{1},D_{2}$ &  & Diameters of primaries \\ 
$d \,\sim 1$ g cm$^{-3}$ &  & Physical density of
bodies \\ 
$a_{\odot} \,\sim 45$ AU &  & Barycenter heliocentric semi-major axis\\ 
$e_{\odot}$ &  & Barycenter heliocentric eccentricity  \\ 
$f_{\odot}$ &  & Barycenter heliocentric  true anomaly  \\ 
$\Omega _{\odot}$ &  & Barycenter heliocentric orbital frequency  \\ 
$R_{H}=a_{\odot}\left( \frac{m_{1}+m_{2}}{3M_{\odot }}\right)
^{\frac{1}{3}}$ &  & 
Radius of binary Hill sphere: $R_{H}=\frac{1}{3^{1/3}}$ in Hill units\\ 
$a_1, a_2$ & & Semi-major axes of primaries \\
$(a,e,i,\tau,\phi,\omega)$ &  & 
Hill orbital elements of binary barycenter \\ 
$b=|a| = |a_2-a_1|$ &  & Impact parameter\\ 
$\Sigma $ &  & Surface mass density of primaries \\ 
$V$ &  & Velocity dispersion of primaries in physical units \\ 
$v$ &  & Velocity dispersion of primaries in Hill units \\ 
$V_{H}=\left[ \frac{G(m_{1}+m_{2})}{R_{H}}\right]
^{\frac{1}{2}}$ &  & Hill
velocity of primaries in physical units\\ 
$v_{H} \,\sim 1.2$ &  & Hill velocity of primaries in Hill units\\ 
$V_{K} \,\sim 4.4$ km/s &  & Keplerian velocity of primaries at 45 AU in physical units \\ 
$v_{K}$ &  & Keplerian velocity of primaries in Hill units \\ 
$\Gamma$ &  & Jacobi constant \\ 
$T_\odot$ & & Orbital period at 45 AU in physical units ($\approx 300$ years)\\
$T = 2 \pi$ & & Orbital period at 45 AU in Hill units\\
$T_{\text{Hill}} $ &  & Hill lifetime; maximum lifetime of an orbit inside the Hill sphere (years)\\
Transient, proto- or quasi-bound binary &  & Temporary binary inhabiting the Hill sphere\\
Nascent binary &  & Newly stabilized binary following a single intruder scattering event\\
Binary hardening (softening) & & A process which increases (decreases) the binary binding energy \\
 &  &  \\ 
&  &  \\ \hline\hline \label{table1}
\end{tabular}
\end{center}

\end{table*}

\section{Three-body Hill approximation}
\label{tbhill}

The general plan of attack is as follows: two individual TNOs are initially 
assumed to be orbiting the Sun on (in general, elliptical) Keplerian orbits proximate to a central Keplerian orbit with semi-major axis $a_{\odot}$. These 
objects may, depending on their relative orbital elements, 
approach to within their mutual Hill sphere radius, $R_H$ - see Table 1 \citep{Murray:1999,Goldreich:2004}. If the two bodies then happen to get caught up in a chaotic layer the resulting quasi-bound binary may be permanently captured by gravitational scattering with an intruder. 

The circular restricted three-body problem (CRTBP) and the three-body Hill problems \citep{Hill:1878,Szebehely:1967,Murray:1999} have proved to be fruitful as test-beds for the study of capture and chaos in dynamical systems \citep{Murray:1999,Simo:2000,Belbruno:2004,Astakhov:2003,Astakhov:2004,Xia:1992}. In its most usual form, the Hill problem consists of two small bodies, $m_1$ and $m_2$, orbiting a third, much larger, body, $m_0$ (hereafter, the Sun, i.e., $m_0=M_{\odot }$) with the center of mass - ``the barycenter'' - of $m_1$ and $m_2$ following a circular orbit. If the barycenter follows an elliptical orbit then the elliptical Hill problem results. Throughout we will refer to the candidate binary partners, $m_1$ and $m_2$, as the ``primaries'' and will assume, without loss of generality, that $m_2 \le m_1$. Later a fourth body - the ``intruder,'' mass $m_3$ - will be introduced which may undergo gravitational scattering with the binary under the governing influence of the Sun.

In Hill's problem at the Hill sphere radius an equilibrium exists between the solar tidal force and the mutual attraction of the two particles. Thus, the Hill sphere radius provides a natural distance scale to describe the motion of particles for which solar tides are a perturbation. Therefore, throughout we use Hill units which ``blow-up'' distances - and orbital elements - in the vicinity of the binary barycenter so that the radius of the Hill sphere, together with distances and velocities are all typically of order unity \citep{Murray:1999}.

Although the circular Hill problem can be derived as a special case of the CRTBP \citep{Szebehely:1967,Murray:1999}, as has been pointed out by H{\'e}non and Petit \citep{Henon:1986,Petit:1986}, the CRTBP and Hill problems are physically distinct. In the CRTBP two masses are assumed to be much larger than the third; the Hill problem emerges if it is further assumed that $m_0 >> m_1 >> m_2$ - this ``hierarchical'' case is shared by the CRTBP and Hill's problem. However, Hill's approximation is more general and is valid for arbitrary values of the mass ratio $m_1/m_2$. Therefore, the Hill approximation is suitable for treating the dynamics of TNBs for which the ratio  $m_1/m_2$ is, in principle, arbitrary and, remarkably, is often of order unity. 
\subsection{Hamiltonian and equations of motion}
\label{ham}
In practice the three-dimensional (spatial) elliptic Hill problem can be derived most directly from the elliptic restricted three-body problem in a similar procedure to the circular case. The elliptic Hill Hamiltonian is the following \citep{Szebehely:1967,Ichtiaroglou:1980,Moons:1988,Llibre:1990,Brumberg:1990,Astakhov:2004,Astakhov:2005,Palacian:2006}:
\begin{eqnarray}
H = E = \frac{1}{2}(p_{\xi}^{2}+p_{\eta}^{2}+p_{\zeta}^{2})\nonumber \\+\frac{1}{2}
(\xi^{2}+\eta^{2}+\zeta^{2}) 
 -(\xi\,p_{\eta}-\eta\,p_{\xi}) \nonumber \\
-\frac{1}{(1+e_\odot\,\cos f_\odot)}\left(
\frac{3\xi^{2}}{2}+\frac{1}{\left| \brho \right|}\right) + 
\frac{81^{\frac{1}{3}}}{2}. \label{eqn:hill}
\end{eqnarray}
\noindent{Here} $E$ is the energy, $(\xi,\eta,\zeta)=\brho$ defines the relative
distance between the binary members $m_{1}$ and
$m_{2}$ and $(p_\xi,p_\eta,p_\zeta)=\mathbf{p}$ is the corresponding momentum vector. The coordinate system $(\xi,\eta,\zeta)$ is rotating with constant angular velocity ${\bf \Omega_\odot}=(0,0,1)$ in the $\xi-\eta$ plane. The eccentricity and true anomaly of the heliocentric orbit of the binary barycenter are $e_\odot$ and $f_\odot$ respectively. In this coordinate system the barycenter is located at the origin.  The additive constant is chosen such that the Lagrange saddle points \citep{Murray:1999} in the circular ($e_\odot = 0$) limit occur at $E=0$. 

Defining the reduced mass $\nu \le 1/2$
\begin{eqnarray}
\nu =\frac{m_2}{m_1+m_2}   \label{eqn:nu}
\end{eqnarray}
\noindent allows the separate motions of $m_1$ and $m_2$ to be recovered
\begin{eqnarray}
\mathbf{\rho}_1=-\nu \thinspace \mathbf{\rho}  \label{eqn:rho1}
\end{eqnarray}
\begin{eqnarray}
\mathbf{\rho}_2=(1-\nu) \thinspace \mathbf{\rho}.   \label{eqn:rho2}
\end{eqnarray}

The equations of motion are the following:
\begin{eqnarray}
\ddot{\xi } &=&2\dot{\eta }+\frac{3\xi
}{\Delta }-\frac{1}{%
\Delta }\frac{\xi }{\rho ^{3}}  \nonumber \\
\ddot{\eta } &=&-2\dot{\xi
}-\frac{1}{\Delta }\frac{\eta }{%
\rho ^{3}}  \nonumber \\
\ddot{\zeta } &=&-\zeta
-\frac{1}{\Delta }\frac{\zeta}{%
\rho ^{3}} \label{eqn:eom}
\end{eqnarray}
where $\Delta = 1/(1+e_\odot\,\cos f_\odot)$ and dots denote derivatives with respect to time. In the circular Hill problem ($e_\odot = 0$) there exists an integral of the motion, the Jacobi constant, $\Gamma$ \citep{Szebehely:1967,Murray:1999}, 
\begin{eqnarray}
\Gamma = 3\xi^2-\zeta^2+\frac{2}{\rho}-(\dot{\xi}^2+\dot{\eta}^2+\dot{\zeta}^2) + 81^{\frac{1}{3}}.\label{eqn:jacobi}
\end{eqnarray}
 \begin{figure*}
\begin{center}$
\begin{array}{cc}
\includegraphics[scale=0.25]{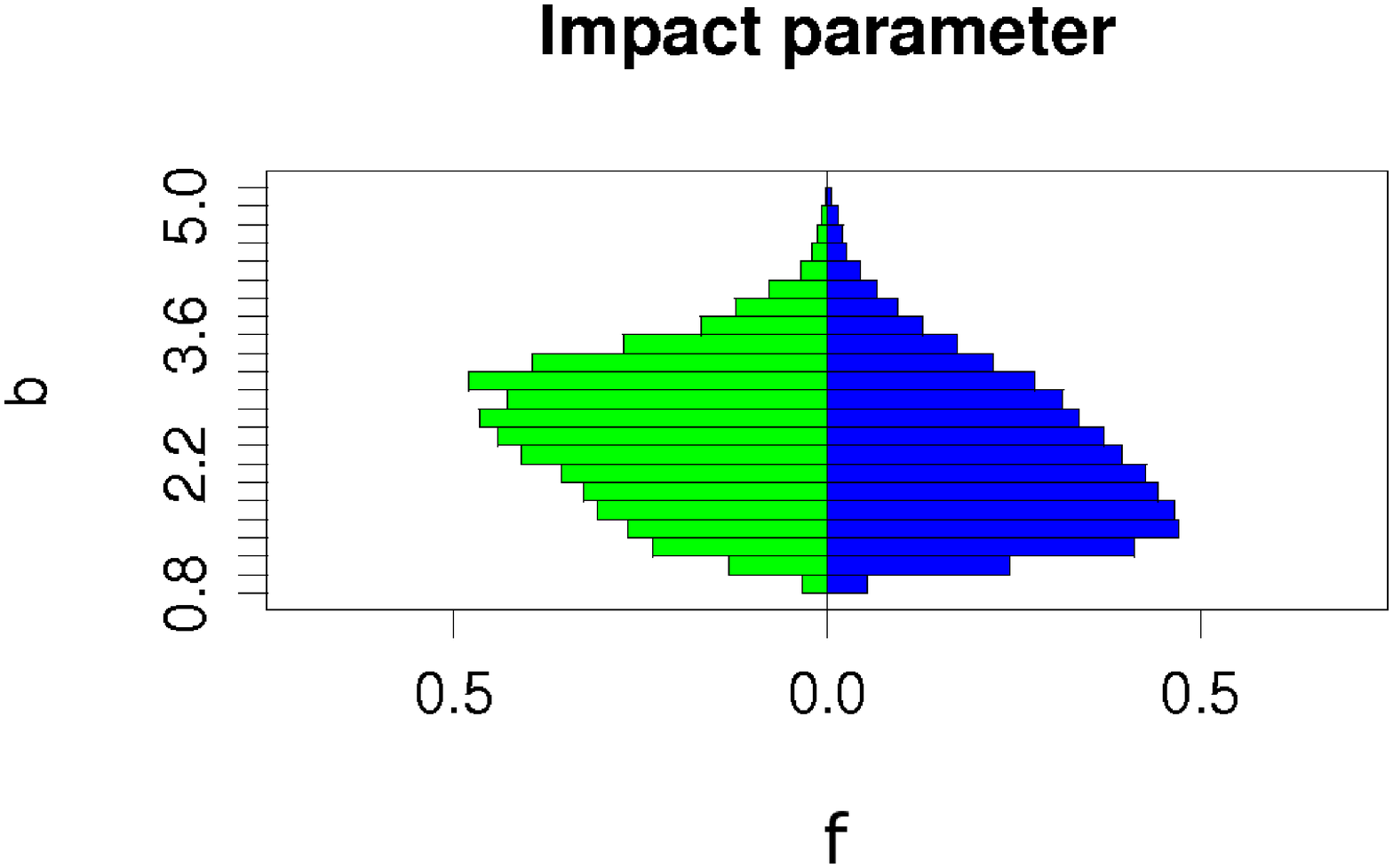}&
\includegraphics[scale=0.25]{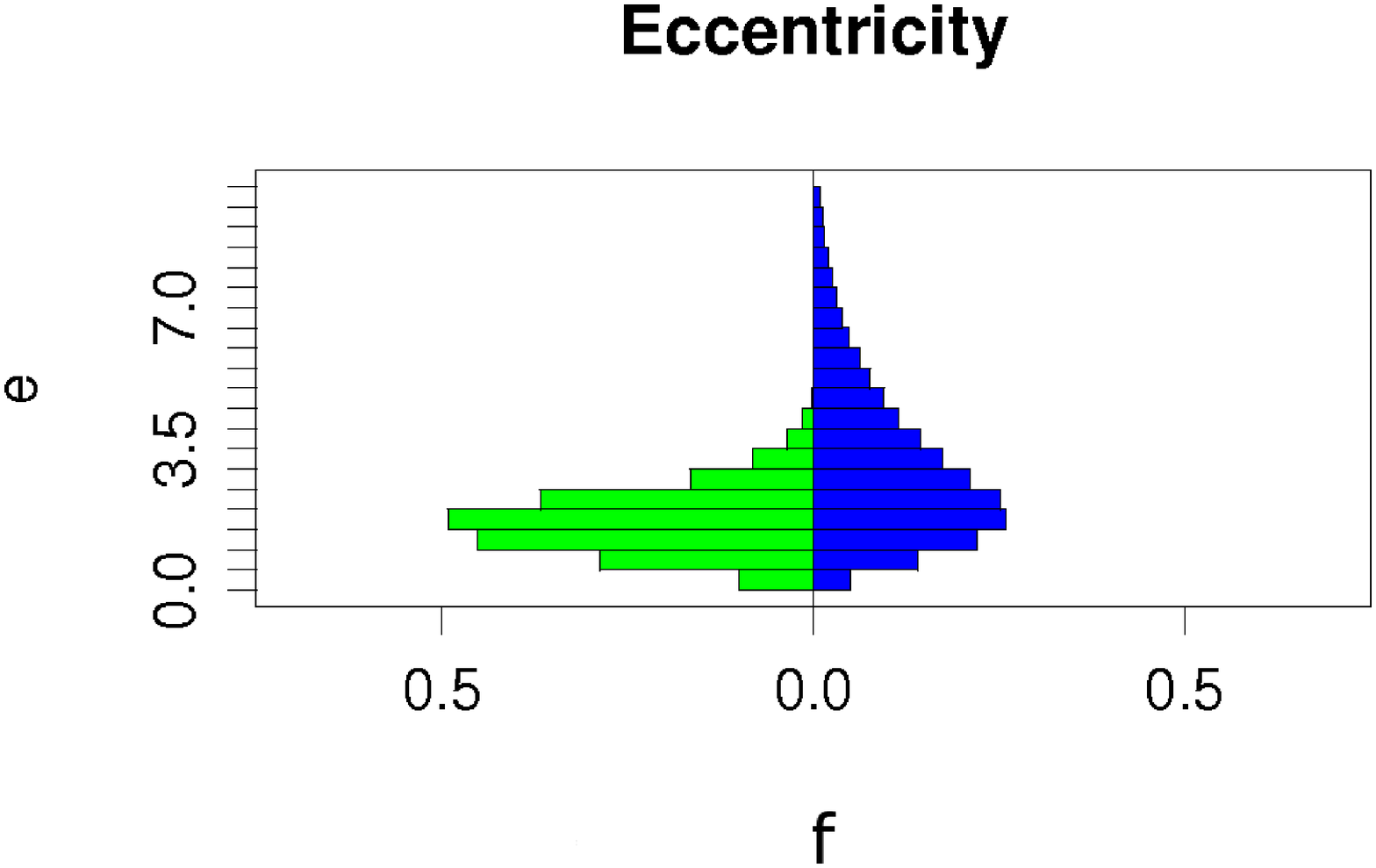}\\
\includegraphics[scale=0.25]{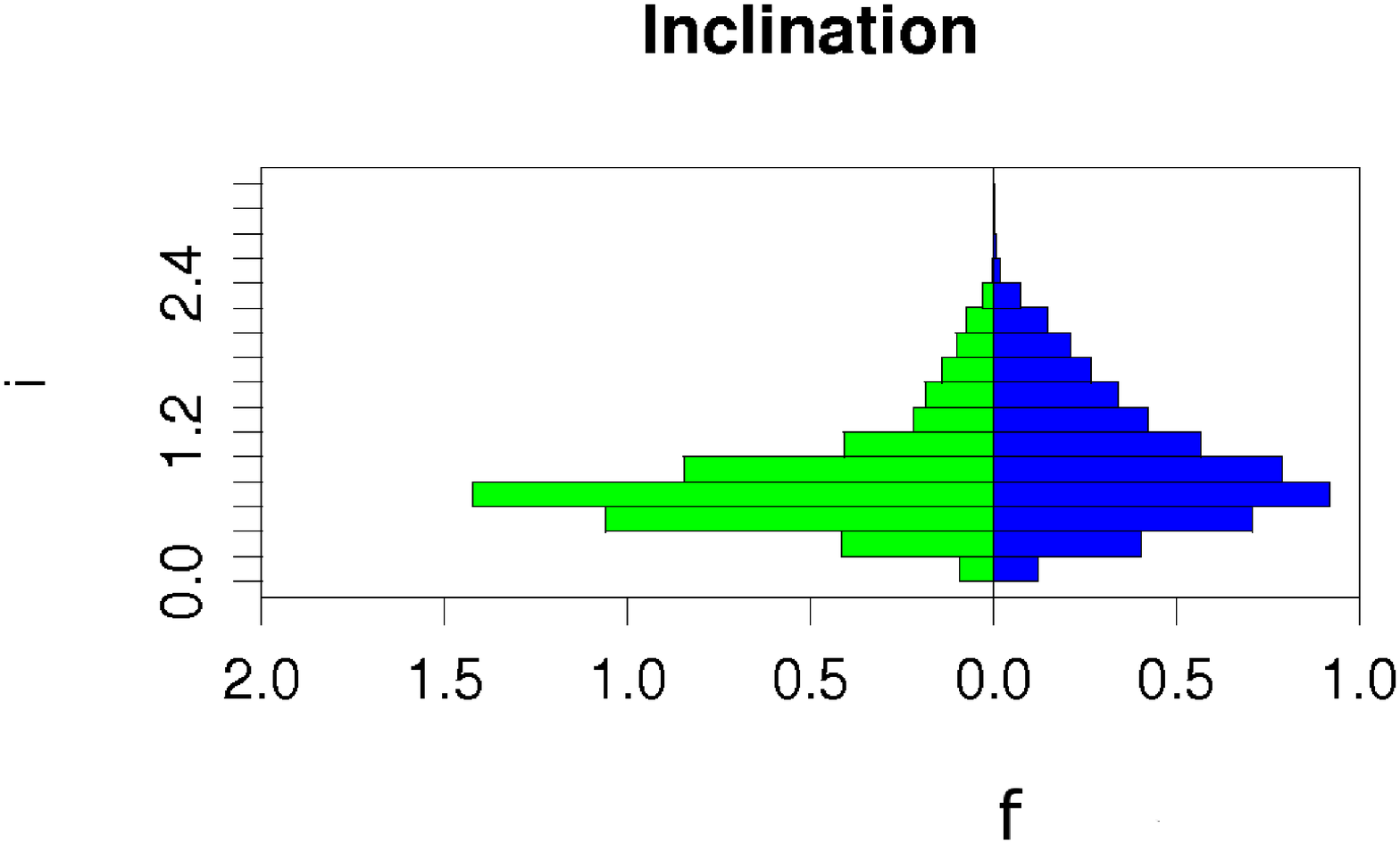}&
\includegraphics[scale=0.25]{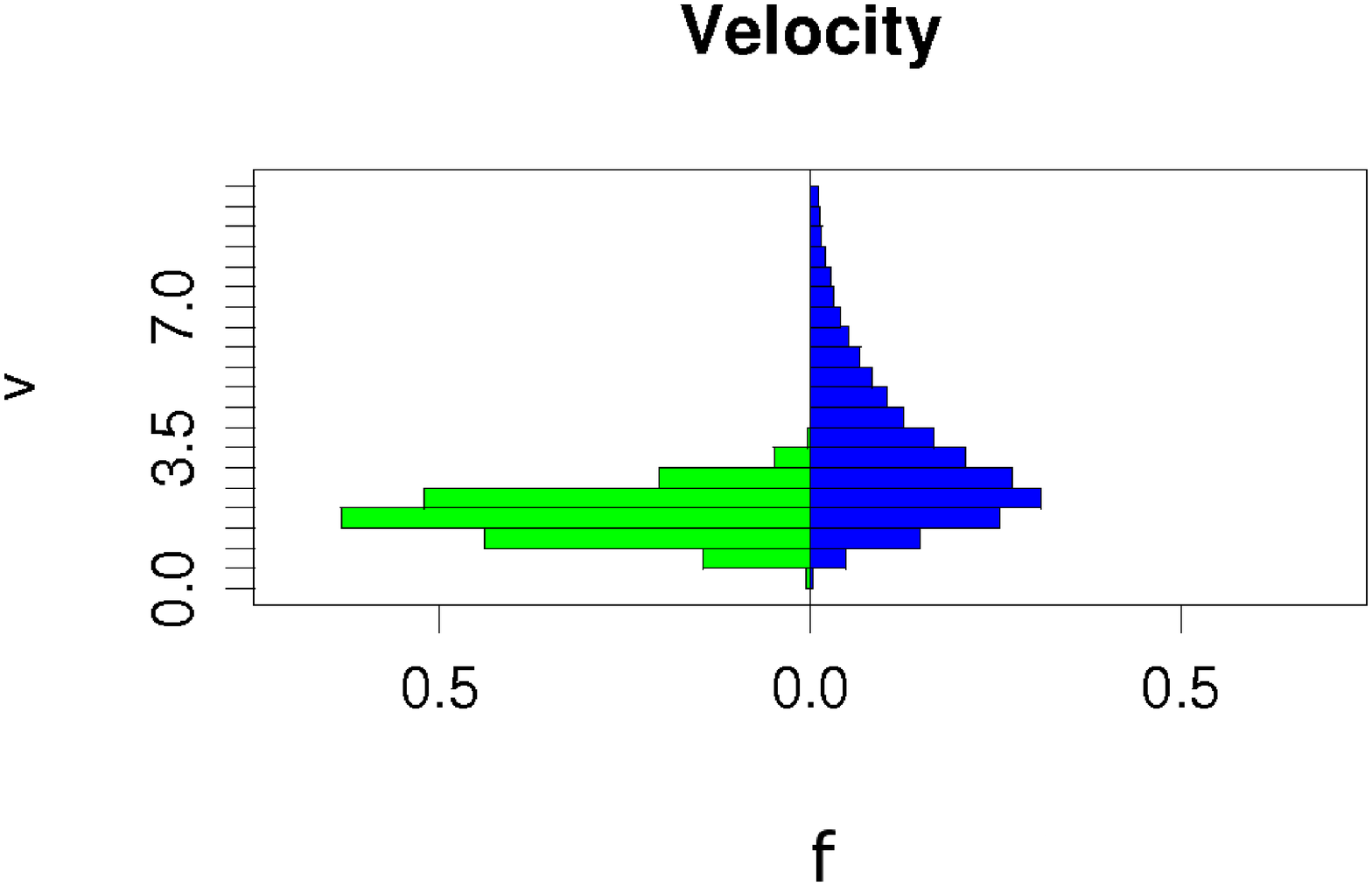}
\end{array}$
\end{center}

\caption{\label{fzone} (colour online) Back-to-back histograms showing the number density of orbits entering the Hill sphere as a function of the parameters, $b$, $e$ and $i$ and the relative velocity $v$ - see eq.\  (\ref{eqn:relv}) - in the circular [green (online), light grey (print)\ and elliptic [blue (online), dark grey (print)] Hill problems. These quantities are all defined in the asymptotic region. Approximately 350,000 trajectories were integrated in both the circular and elliptic cases. The area under each histogram is normalized to unity; $f$ is the probability density. All units are Hill units. The Jacobi constant is picked randomly as described in the text. }
\end{figure*}

\subsection{Chaos in the Hill sphere}
If two bodies come within their mutual Hill sphere they cannot be captured into a binary permanently unless an energy loss mechanism is present. However, the partners can form a transient (quasi-bound) binary which may have a substantial lifetime \citep{Astakhov:2005}. In order to form a quasi-bound binary the two primaries
must first come close enough that mutual gravity dominates solar differential gravity. Thus, the proto-binary effectively defines a Hill sphere.  At low energies the gateways to this region are the Lagrange saddle points, $L_1$ and $L_2$, which separate interior and exterior (heliocentric) orbits. The interior region is the Hill sphere and has radius
\begin{eqnarray}
R_{H}=a_{\odot}\left( \frac{m_{1}+m_{2}}{3M_{\odot }}\right)
^{\frac{1}{3}}. \label{eqn:hsphere}
\end{eqnarray}

Examination of Poincar\'{e} surfaces of section \citep{Lichtenberg:1992} in the Hill problem \citep{Simo:2000,Astakhov:2003,Astakhov:2004,Astakhov:2005} -- or, equivalently, the CRTBP for small masses \citep{Astakhov:2003} -- reveals that, even at energies {\it above} the Lagrange saddle points, $L_1$ and $L_2$, phase space is divided into regular Kolomogorov-Arnold-Moser (KAM) regions \citep{Lichtenberg:1992}, chaotic regions and hyperbolic (direct scattering) regions \citep{Zaslavsky:1985}. Most importantly, the chaotic orbits separate the regular from the hyperbolic regions. Because incoming trajectories from outside the Hill sphere cannot penetrate the regular KAM tori at all in 2-dimensions (2D) and enter regular regions exponentially slowly in 3D \citep{Nekhoroshev:1977,Perry:1994,Lichtenberg:1992,Astakhov:2003}, particles entering the Hill sphere from heliocentric regions must either enter chaotic layers or scatter out of the Hill sphere promptly. Those that enter chaotic layers may remain temporarily trapped within the Hill sphere on time scales that are orders of magnitude larger than might be expected from estimates based on prompt transit of the Hill sphere. This dramatically increases their chances of being permanently captured by, e.g., gravitational scattering with a fourth body.

It is interesting to note that chaos in the Hill sphere is similar to the situation in several problems of current interest in atomic and molecular physics; these include the creation of non-dispersive electronic Trojan wave packets in Rydberg atoms \citep{Bialynicki-Birula:1995,Farrelly:1995,Lee:1995,Lee:1997,Lee:2000,Kalinski:2005}; the dynamics and ionization of Rydberg atoms and molecules in external microwave \citep{Griffiths:1992,Farrelly:1994,Farrelly:1995,Farrelly:1995a,Deprit:1996,Brunello:1997,Bellomo:1997} or magnetic fields \citep{Johnson:1983,Saini:1987,Uzer:1991}; and the dynamics of ions in ion traps \citep{Howard:1993,Elipe:2002}. 
\subsection{Penetration of the mutual Hill sphere by the primaries}

The problem is similar to that studied by \citet{Henon:1986} who investigated satellite encounters in the planar circular Hill problem - see also \citet{Yoder:1983,Petit:1986}: The two primaries are initially assumed to follow elliptical heliocentric orbits with semi-major axes $a_1$ and $a_2$ with some velocity dispersion, $V$, around an elliptical Keplerian orbit lying in the invariant plane with semi-major axis $a_{\odot}$. Assume that the Keplerian velocity, $V_K$, is much greater than the relative velocity of the two primaries as is thought to have been the case in the primordial Kuiper-belt, i.e., prior to dynamical excitation \citep{Chiang:2006,Levison:2006,Gladman:2006,Quillen:2004}: How do the values of their orbital elements ``at infinity'' determine how close the primaries will approach and, if they do so approach, then is the encounter favourable for binary production? To answer these questions we first need to consider the asymptotic pre-encounter behaviour of the system.

\subsubsection{Hill orbital elements}
The asymptotic behaviour, when the relative separation between the primaries, $\rho$, is large ($\left| t \right| \rightarrow \infty$), is complicated by the infinite range nature of the ``$1/\rho$'' interaction potential. Integrations originating in the asymptotic regime must start at some finite $t = t_0 < \infty$. In principle, one could start the integrations at extremely large separations such that the mutual interaction terms in eqs.\ (\ref{eqn:eom}) are sufficiently small. Alternatively, one can use asymptotic expansions to start the incoming solution (from infinity) and to continue the outgoing solution (to infinity). In between the solution is obtained numerically by integrating the equations of motion. High-order asymptotic solutions have been derived by H{\'e}non and Petit in the planar circular Hill problem \citep{Henon:1986,Petit:1986} and by \citet{Brumberg:1990} in the three-dimensional elliptic Hill problem. The asymptotic solutions are, to lowest order,
\begin{eqnarray}
\xi &=&a-e\thinspace \cos (t-\tau) \nonumber \\
\eta &=&-\frac{3}{2} \thinspace a\thinspace(t-\phi) + 2\thinspace e\thinspace  \sin (t-\tau) \nonumber \\
\zeta &=&i\thinspace \sin (t-\omega) \nonumber \\
\dot{\xi} &=&e \thinspace \sin (t-\tau) \nonumber \\
\dot{\eta} &=&-\frac{3}{2} \thinspace a + 2\thinspace e\thinspace \cos (t-\tau) \nonumber \\
\dot{\zeta} &=&i\thinspace \cos (t-\omega) \label{eqn:asymp}
\end{eqnarray}
where $a, e$, and $i$ are sometimes called the Hill, or ``reduced,''  orbital elements: here  $b = |a|$ is the impact parameter, $e$ is the eccentricity and $i$ is the inclination in Hill units while $\tau, \phi$, and $\omega$ are phase angles. Alternatively, $b$ can be thought of as the fractional distance of the semi-major axis from the reference orbit $a_\odot$. Explicitly the reduced elements are related to the usual semi-major axis ($a_{c}$), eccentricity ($e_{c}$) and inclination ($i_{c}$) in the CRTBP as follows \citep{Greenzweig:1990}.

\begin{eqnarray}
a &=& \frac{(a_{c}-1)}{R_H} \nonumber \\
e &=& \frac{e_{c}}{R_H}  \nonumber \\
i &=& \frac{\sin (i_{c})}{R_H}  
\end{eqnarray}

The set of Hill orbital parameters has been used extensively in studies of the accretion of planetesimals or the dynamics of particles in planetary rings \citep{Nakazawa:1988,Nakazawa:1989,Ohtsuki:1990,Wetherill:1989,Greenzweig:1990,Greenzweig:1992,Wetherill:1993,Ohtsuki:2002,Stewart:2000,Petit:1987,Petit:1987a,Rafikov:2001,Rafikov:2003,Rafikov:2003a,Rafikov:2003b,Rafikov:2003c}. 

In the circular limit the quantity $\phi$ can be eliminated by an appropriate choice of the origin of time resulting in five elements. Equations (\ref{eqn:asymp}) are for the relative motion of the primaries; by attaching subscripts, $j=1,2$, one may obtain corresponding expressions for the individual primary orbits. 
\begin{figure*}
\begin{center}$
\begin{array}{cc}
\includegraphics[scale=0.25,angle=270]{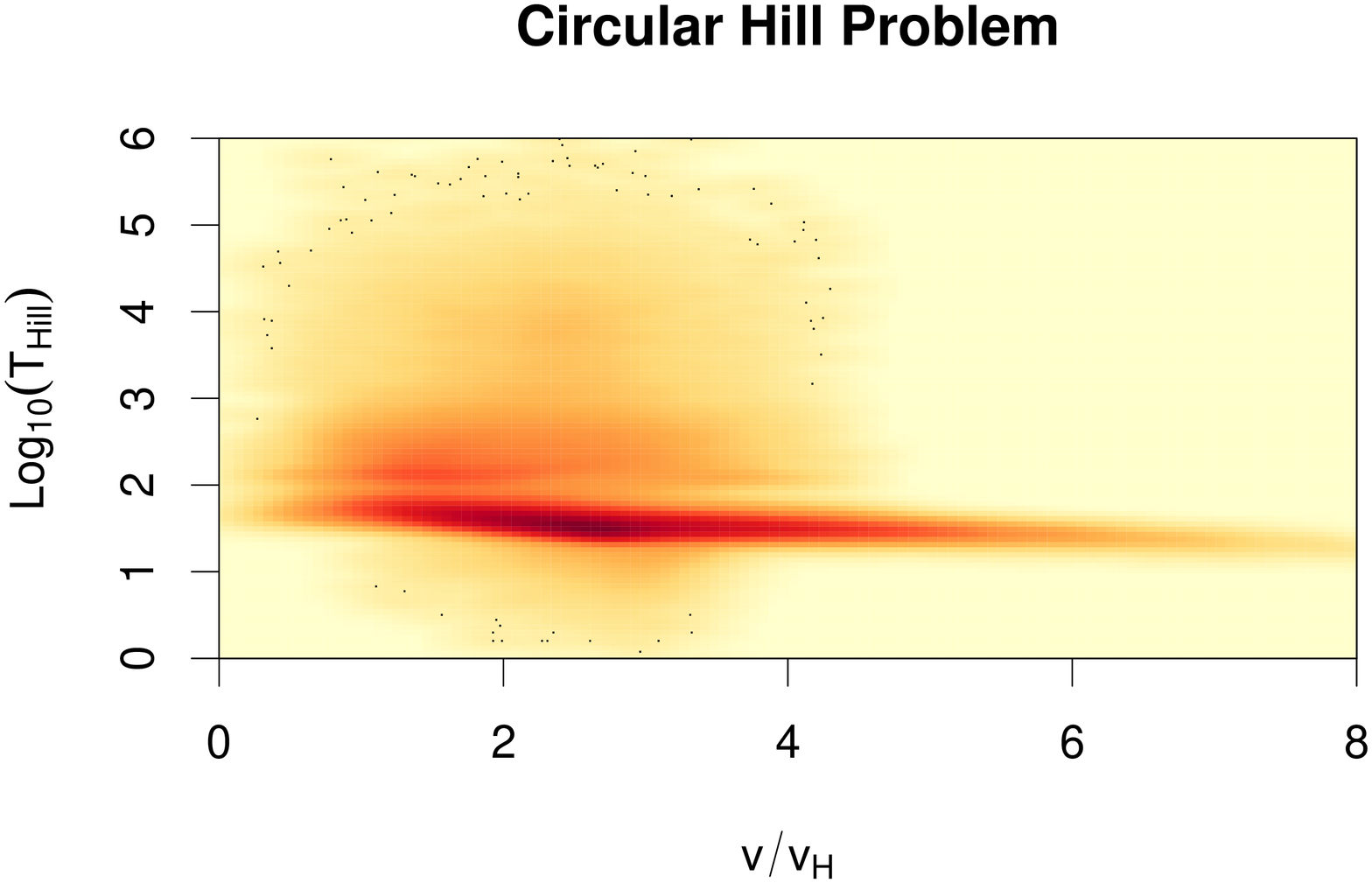}&
\includegraphics[scale=0.25,angle=270]{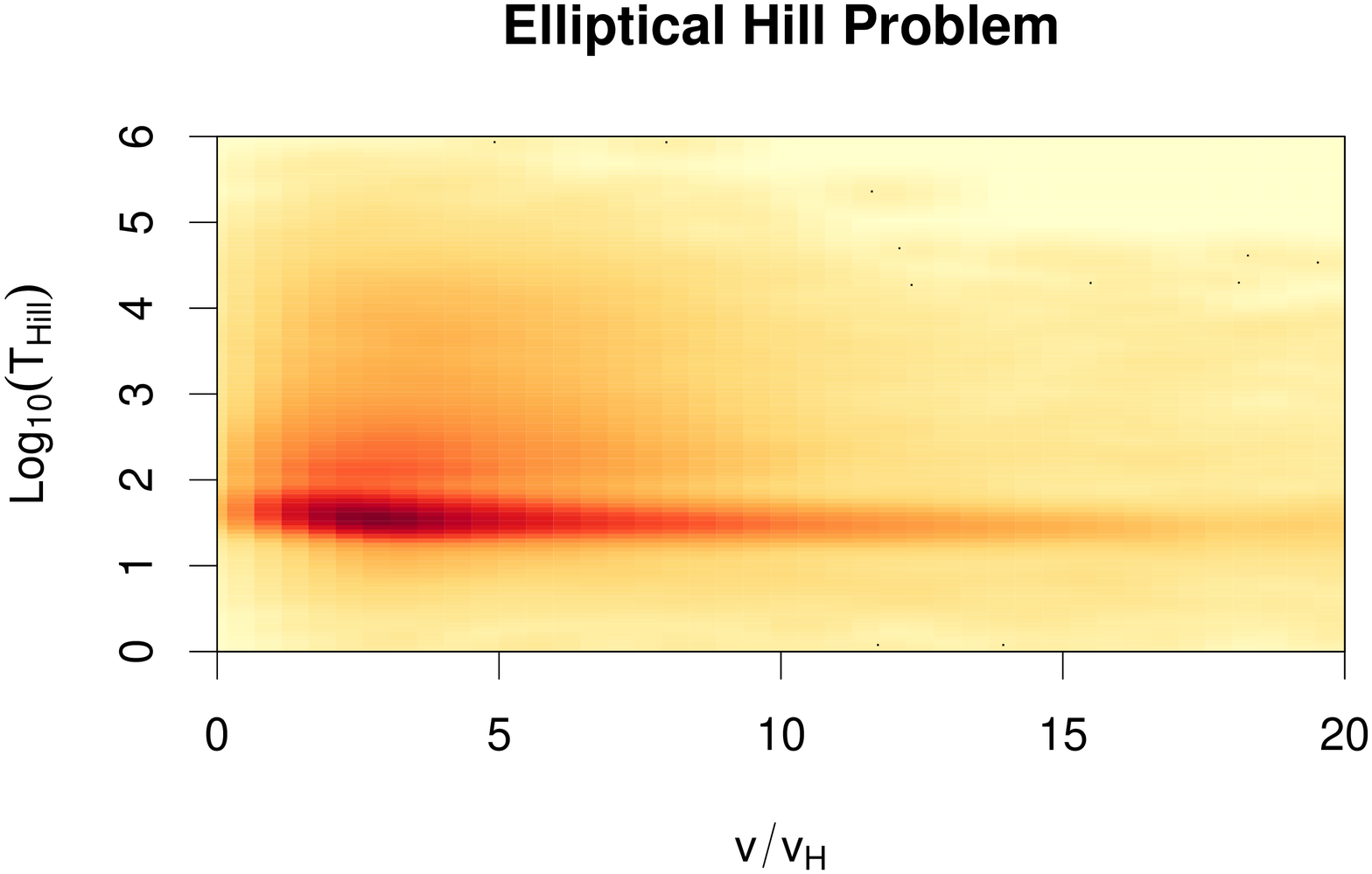}
\end{array}$
\end{center}

\caption{\label{ltime} (colour online) Kernel smoothed scatterplots showing $\log_{10}(T_{\text{Hill}})$ where $ T_{\text{Hill}}$ is the Hill lifetime (in years) in the circular (left) and elliptic (right) cases as a function of the relative asymptotic velocity $v$ scaled by $v_H$ - see eq.\ (\ref{eqn:relv}).  In Hill units 300 years $\sim 2 \pi$; i.e., the Hlll unit of time $\sim 50$ years. Scale runs from low [yellow (online), light grey (print)] to high [red (online), dark grey (print)]. In this representation smoothed point densities are computed using a kernel density estimator. In regions of sparse data the plot is augmented by black dots which represent individual data points \citep{Gentleman:2006,Carr:1987,gene}. For further discussion of kernel smoothing see Appendix A.}
\end{figure*}

The Jacobi constant can be expressed in closed form in terms of the set $a,e,i$
\begin{eqnarray}
\Gamma = \frac{3}{4}a^2-(e^2+i^2) + 81^{\frac{1}{3}}.\label{jac1}
\end{eqnarray}
For our purposes it is sufficient to use the lowest-order versions of  the asymptotic expansions - eq. (\ref{eqn:asymp}) - which correspond to Keplerian motion and which are valid in both the pre- and post-encounter asymptotic regions. Strictly, higher order corrections should be included; however, even if the interaction is not negligible the orbit can, instantaneously, still be described by elements having the form of eq.\ (\ref{eqn:asymp}). Assume that, in this way, we compute a set of orbital elements $(a^\prime,e^\prime,i^\prime)$ which are approximations to the true orbital elements $(a,e,i)$. Provided that the error is small then one can think of $(a^\prime,e^\prime,i^\prime)$ as being the exact orbital elements but for a slightly different value of the Jacobi constant, $\Gamma^\prime$. In our Monte Carlo simulations $\Gamma$ is chosen randomly and, therefore, neglecting to use the higher order asymptotic formulae is not expected to affect the results. The main reason for not using higher-order formulae is the additional computational effort involved in solving for the orbital elements, which must be done iteratively. 
\subsubsection{The Hill velocity and the relative velocity}
Physically, the Hill velocity is the orbital velocity around a large body at the Hill sphere radius assuming no solar perturbations \citep{Goldreich:2004,Rafikov:2003c,Murray-Clay:2006} - so that one revolution at the Hill radius completes in one orbital
period around the Sun. However, this turns out to be a fairly approximate quantity because the dynamics is essentially three-body in the case of TNB formation. We will modify this definition slightly to include the case where the Hill sphere is defined by a binary, rather than by a single mass, but for which the separation between partners is considerably less than $R_H$, 
\begin{equation}
V_{H}=\left[ \frac{G(m_{1}+m_{2})}{R_{H}}\right]
^{\frac{1}{2}} \sim \Omega_\odot \thinspace R_H. \label{hillv}
\end{equation}
In Hill units, the Hill velocity $v_H = 3^{-1/6} \sim 1.2$. To see the correspondence in physical units, consider a binary with the following characteristics \citep{Veillet:2002}: (a) a barycenter semi-major axis of 45 AU; (b) 100 km radii binary partners; and, (c) density $d$ = 1 g cm$^{-3}$. For these parameters $R_H \sim 7.5 \times 10^5$ km, $m_1 = m_2 \sim 4.0 \times 10^{18}$g, the Keplerian velocity, $V_K \sim 4.4$ km/s and the Hill velocity $V_H \sim 0.9$ m/s. 

Scattering is said to be in the shear-dominated regime (dynamically cold) when $v \lesssim v_H$ and in the dispersion-dominated regime (dynamically hot) if  $v \gtrsim v_H$ \citep{Stewart:2000,Collins:2006}. Shear is induced by the difference between the Keplerian angular velocities of primaries having different heliocentric semi-major axes.  If the relative approach velocity of two particles is greater than the differential shear in the disk across the Hill (or tidal) radius then the dynamics is dispersion-dominated \citep{Goldreich:2004,Rafikov:2003b} The parameter which determines whether the system is hot or cold is the ratio of the velocity dispersion to the shear across the Hill radius, i.e., $v/\Omega_\odot \thinspace R_H= v/v_H$ \citep{Rafikov:2001}. Thus, the Hill velocity serves roughly to demarcate the transition from two- to three-body dynamics for encounters between ``small''  ($R \lesssim  1$ km) and ``big''  ($R \gtrsim 100$ km) bodies; i.e., if the relative speed of bodies undergoing close encounters is greater than the Hill velocity then two-body dynamics is expected to provide a good approximation to the dynamics otherwise solar tides must be included, i.e., three-body effects are important \citep{Rafikov:2003c,Goldreich:2004}. However, this is not a sharp transition and \citet{Rafikov:2003c} has identified a transition regime. In this work we find that binary capture spans such a transition region.

\citet{Rafikov:2003c} has further argued that the growth of big bodies (planetary embryos) in a planetesimal disk undergoes a direct transition from a relatively short runaway growth phase to a much longer oligarchic phase. If TNBs formed during the longer oligarchic phase then this would suggest that the velocity dispersion was larger than $v_H$. In contrast, \citet{Goldreich:2002} have estimated that the velocities of $\sim 100$ km sized bodies were on the order of $v_H/3$ in the early Kuiper-belt. As pointed out by \citet{Chiang:2006} if the velocity dispersion, $v$, of big bodies is less than $v_H$ then these bodies will collapse into an essentially two-dimensional disk due to so-called runaway cooling induced by the dynamical friction exerted by the sea of small bodies. Therefore, TNB mutual orbit inclinations ought to be similar, i.e., mutual orbit normals will be approximately parallel. In fact, TNB inclinations appear to be randomly distributed \citep{Noll:2003,Chiang:2006}. This implies an isotropic primordial velocity distribution and suggests that the big bodies did not all originally lie in the same plane \citep{Noll:2007} and that TNBs formed during the oligarchic phase. 

While these arguments lead to the inference that, in the early Kuiper-belt, $v > v_H$ and that observed TNB inclinations are primordial, it could also be the case that the observed TNB inclinations are the result of either (i) dynamical stirring of the Kuiper-belt after binaries formed \citep{Chiang:2006,Levison:2006} or (ii) post-capture binary hardening \citep{Heggie:1996,Heggie:2003}, or both. In our previous simulations we found that inclinations do not change significantly during the hardening process \citep{Astakhov:2005} - this is similar to the approximate conservation of irregular satellite inclinations undergoing CAC at Jupiter or Saturn \citep{Astakhov:2003}.

A related consideration is deciding how the ``relative velocity'' is best defined. As the two bodies approach each other from infinity then their relative velocity changes and, if the bodies enter a chaotic layer within the Hill sphere, fluctuations in the relative velocity can be large. At large separations we adopt the following definition,
\begin{eqnarray}
v = \frac{R_H}{a_\odot} \sqrt{e^2+i^2}\thinspace v_K
 \label{eqn:relv}
\end{eqnarray}
This expression is similar to the average velocity of a planetesimal, relative to other planetesimals in a swarm with mean Keplerian velocity $v_K$, and as obtained by averaging over an epicycle and a vertical oscillation \citep{Greenzweig:1990,Greenzweig:1992}.  It turns out that the relative velocities at infinity which lead to penetration of the Hill sphere - see Fig. \ref{fzone} - are only roughly on the order of the Hill velocity.
\subsection{Simulations of primary encounters}
The most straightforward approach might appear to be a Monte Carlo simulation in which the equations of relative motion - eq.\ (\ref{eqn:eom}) - are integrated inwards from large initial separations. However, this is not practical computationally because most initial conditions picked randomly at  infinity will not lead to primary encounters, i.e., separations less than $R_H$.  Especially is this true for the three-dimensional elliptical problem. Instead we adopt a different approach which involves the several stages now to be described.

\subsubsection{The feeding zone}
The first task is to try to establish what ranges of initial conditions at infinity can result in penetration of the Hill sphere. If these ranges can be pinned down they define a feeding zone in phase space. Assume that the edges of the presumed zone can be approximately delimited by maximum and minimum values of the three orbital elements. The most obvious constraint is that the Jacobi constant should have a value higher than its value at the Lagrange points, i.e., $\Gamma > 0$. 
We desire to obtain, in addition, constraints on the ranges of each of the orbital elements if such constraints exist. Of course, the mere fact that a set of orbital elements is contained within the feeding zone need not imply that the corresponding trajectory will actually enter the Hill sphere. It is also possible that initial conditions lying outside such a feeding zone will enter the Hill sphere. The feeding zone is, therefore, an approximate, but potentially useful, concept because it narrows down the ranges of the orbital elements which can, in principle, lead to long-lived encounters within the Hill sphere. Reasonably good constraints on the feeding zone can be found numerically as follows.
\begin{enumerate}
\item Generate initial conditions randomly inside the Hill sphere for values of $\Gamma$ chosen randomly and uniformly in the range $\Gamma \in{(0,8})$ in Hill units. 

\item Cartesian positions and velocities are then generated uniformly and randomly inside a sphere of radius $1.2 R_H$. 

\item Next Hill's equations are integrated until the trajectory penetrates the surface of a sphere, $S_2$, in the asymptotic region; here $S_2$ is chosen to have radius $\sim 350 R_H$. At this point the integration is stopped and the orbital elements at infinity stored.

\item Some initial conditions will, of course, have been started within (and, in three-dimensions, between \citep{Nekhoroshev:1977}) regular KAM zones inside the Hill sphere. These initial conditions are of no interest for present purposes because they lie in regions that cannot be penetrated at all in two-dimensions and only exponentially slowly in higher dimensions. Therefore, they are discarded; however, first they must be identified. This is accomplished as follows: if, after a sufficiently long time, $T_{cut}$, a trajectory has not passed through $S_2$ then that orbit is discarded. Of course, it is possible that discarded trajectories did not actually start inside KAM regions but, instead, were trapped within extremely long-lived chaotic zones - i.e., they are amongst the initial conditions of greatest interest. Therefore it is important that $T_{cut}$ be chosen large enough that (a) such cases are relatively few and (b) the  results are insensitive to its precise value. We chose $T_{cut} = 1000$  in dimensionless Hill units or $\sim 50,000$ years. While this is shorter than the very longest lifetimes shown in Fig.\ \ref{ltime} this method can still detect such orbits because it is unlikely, in this approach, that initial conditions selected at random inside the Hill sphere will subsequently spend their entire Hill lifetime inside this region.

\item Because we integrate a very large number of trajectories we are confident that this approach allows for the harvesting of essentially all types of initial condition which, coming in from infinity, will penetrate the Hill sphere. 

\end{enumerate}

Figure \ref{fzone} compares histograms of the asymptotic distributions of orbital elements and the velocity of initial conditions, which, started in the asymptotic regime, go on to penetrate the Hill sphere in the circular and elliptical problems. The distributions in the circular and elliptical cases are generally similar except that the elliptical distributions are fatter; this is most marked in the distribution of mutual orbit eccentricity and velocity. In these simulations the heliocentric eccentricity was confined to the range $e_\odot \in (0,0.3)$. 

Roughly speaking, in the circular Hill problem, the feeding zone is defined by the following ranges (in Hill units): $\Gamma \in (0,4), b \in (0.8,5),  e \in (0,5), \left| i \right| \in (0,1.5)$ and $v \in (0,4)$. These ranges are approximate and initial conditions lying outside them may lead to trajectories which penetrate the Hill sphere but, we will find, they tend to do so promptly, that is they do not get caught up in chaotic layers. Similarly, not all initial conditions lying inside these ranges necessarily pass through - or close to - the Hill sphere. 

\subsubsection{Quasi-binary Hill lifetimes}

We define the Hill lifetime, $T_{\text{Hill}}$, for each set of initial conditions at infinity to be the total time the resulting trajectory spends within the Hill sphere - see Table 1. To calculate $T_{\text{Hill}}$ the trajectories used in identifying the feeding-zone are back-integrated from infinity and their time inside the Hill sphere recorded.

The key to the CAC scenario is the dramatic extension in the time the binary partners spend within the Hill sphere due to their having become entangled in very long lived - though chaotic - regions of phase space. We computed lifetimes of all trajectories which, starting in the asymptotic region, end up penetrating the Hill sphere. The results are shown in Fig.\ \ref{ltime} where Hill lifetimes are plotted as a function of the relative velocity at infinity obtained using eq.\  (\ref{eqn:relv}). The large island lying between $\approx 10$ and $\approx 100$ years corresponds to essentially direct or ``ballistic'' transit through the Hill sphere and appears in both the circular and elliptic cases. In both cases a second large island exists and corresponds to lifetimes on the order of thousands of years or more with much longer lifetimes also being possible. 

Figure \ref{ltime} can be thought of as a nonlinear map which relates asymptotic velocity to transit time through the Hill sphere. Clearly, two (or more) trajectories with the same asymptotic velocity can spend very different times inside the Hill sphere. This map also shows that very small asymptotic velocities tend either not to enter the Hill sphere at all, or, if they do enter, they transit rather quickly. In part, this is because the actual velocity after entering the Hill sphere can be considerably greater than the asymptotic value. Note that Fig.\ \ref{ltime} covers essentially the full spectrum of asymptotic velocities which can lead to Hill sphere penetration.

The elliptic case shown in Fig.\ \ref{ltime}(b) involves a much larger range of asymptotic velocities and orbital elements than does the circular problem.  Simulating the four-body dynamics - i.e., intruder - binary scattering inside the Hill sphere - in the elliptical case will be significantly more demanding computationally than in the circular case because of the larger space of intruder and binary initial conditions, e.g., both the primaries and the intruder are now allowed to follow elliptical orbits. Mainly for this reason, from now on, in this paper, we specialize to the circular Hill problem but consider the elliptic problem worthy of further study.

\section{Stabilization by intruder scattering in the circular Hill problem}
\label{stab4}

Having formed a transient binary the next step is to capture it permanently. In this section we investigate how the masses and velocities of the primaries and the intruder affect nascent (i.e., just-captured) binary formation as well as the properties of any binaries that result. The simulations are done in the four-body Hill approximation \citep{Scheeres:1998} as is now described.

\subsection{Four-body Hill approximation}
Three comparatively small bodies, with a mutual
centre-of-mass, ${\bf R}_c$, orbit a much larger body - the
Sun -  $m_0=1$ on a near-circular orbit {\it e.g.}, the primaries and the intruder. The total mass of the three bodies is defined by

\begin{equation}
\mu =\sum_{j=1}^3 m_j \ll 1  \label{mu}
\end{equation}

\noindent where ${\bf R}_c \approx {\bf a}=(1,0,0)$ defines the
motion of the three-body centre-of-mass along an almost circular
orbit which defines the rotating frame. The vector equations of motion are \citep{Scheeres:1998}

\[
\ddot{\brho} + {\bf\Omega} \times [2 \dot{\brho} + {\bf\Omega}
\times \brho]=-\brho+3{\bf a}({\bf a} \cdot \brho) - ({\alpha}_1 +
{\alpha}_2) \frac {\brho} {|\brho|^3} 
\]

\begin{equation}
+ {\alpha}_3 \left( \frac {{\brho}_3 -{\brho}_2}{|{\brho}_3 -
{\brho}_2|^3} -  \frac {{\brho}_3 -{\brho}_1}{|{\brho}_3 -
{\brho}_1|^3}\right)
  \label{rho}
\end{equation}

\[
\ddot{\brho}_3 + {\bf\Omega} \times [2 \dot{\brho}_3 + {\bf\Omega}
\times {\brho}_3]=-{\brho}_3+3{\bf a}({\bf a} \cdot {\brho}_3) -
\]

\begin{equation}
 {\alpha}_1  \frac {{\brho}_3 -{\brho}_1}{|{\brho}_3 -
{\brho}_1|^3} -  {\alpha}_2 \frac {{\brho}_3
-{\brho}_2}{|{\brho}_3 - {\brho}_2|^3}
   \label{rho3}
\end{equation}

Here ${\brho}_3$ is the coordinate of the third intruder body, $m_3$, and
$m_{j}=\mu {\alpha }_{j}$ where

\begin{equation}
\sum_{j=1}^3 {\alpha}_j = 1.  \label{alpha}
\end{equation}

When $m_{3}=0$ eq. (\ref{rho}) reduces to the three-body Hill
problem \citep{Hill:1878,Szebehely:1967,Murray:1999} and becomes uncoupled from
eq. (\ref{rho3}). 

\subsection {Numerical procedure}

This numerical simulations were performed as follows.
\begin{enumerate}
\item Initial conditions for the primaries ``at infinity,'' and which are guaranteed to penetrate their mutual Hill sphere, are generated as described in Subsec. 2.4. That is, initial conditions are generated randomly inside the Hill sphere and integrated until the trajectory penetrated a sphere, $S_2$, of radius $\sim 350 R_H$ at which point the integration is stopped and the orbital elements computed.

\item The integration is then run backwards so as to compute the Hill lifetime of the orbit and the orbital elements are stored. 

\item This procedure is repeated until a sufficiently large cohort of initial conditions at infinity has been generated.

\item Intruder initial conditions at infinity are generated in a similar way except that (a) initial conditions for the outward integration are generated inside a sphere of radius $2R_H$ and (b) the radius of the sphere ``at infinity'' was chosen randomly in the range $30R_H < S_2 < 350 R_H$. This is done to allow for the possibility of stabilization by ``near-misses'', i.e., intruders which do not actually penetrate $R_H$ and also to ensure that the phases of the intruders with respect to the binary orbit are varied.

\item Next, the primaries are integrated back from infinity in the three-body Hill approximation until they come within a radius $R_H < R_0 < 5R_H$ of each other. The actual radius, $R_0$, is generated randomly and the orbital parameters of the primaries are then stored.
\begin{figure*}
\begin{center}$
\begin{array}{cc}
\includegraphics[scale=0.5]{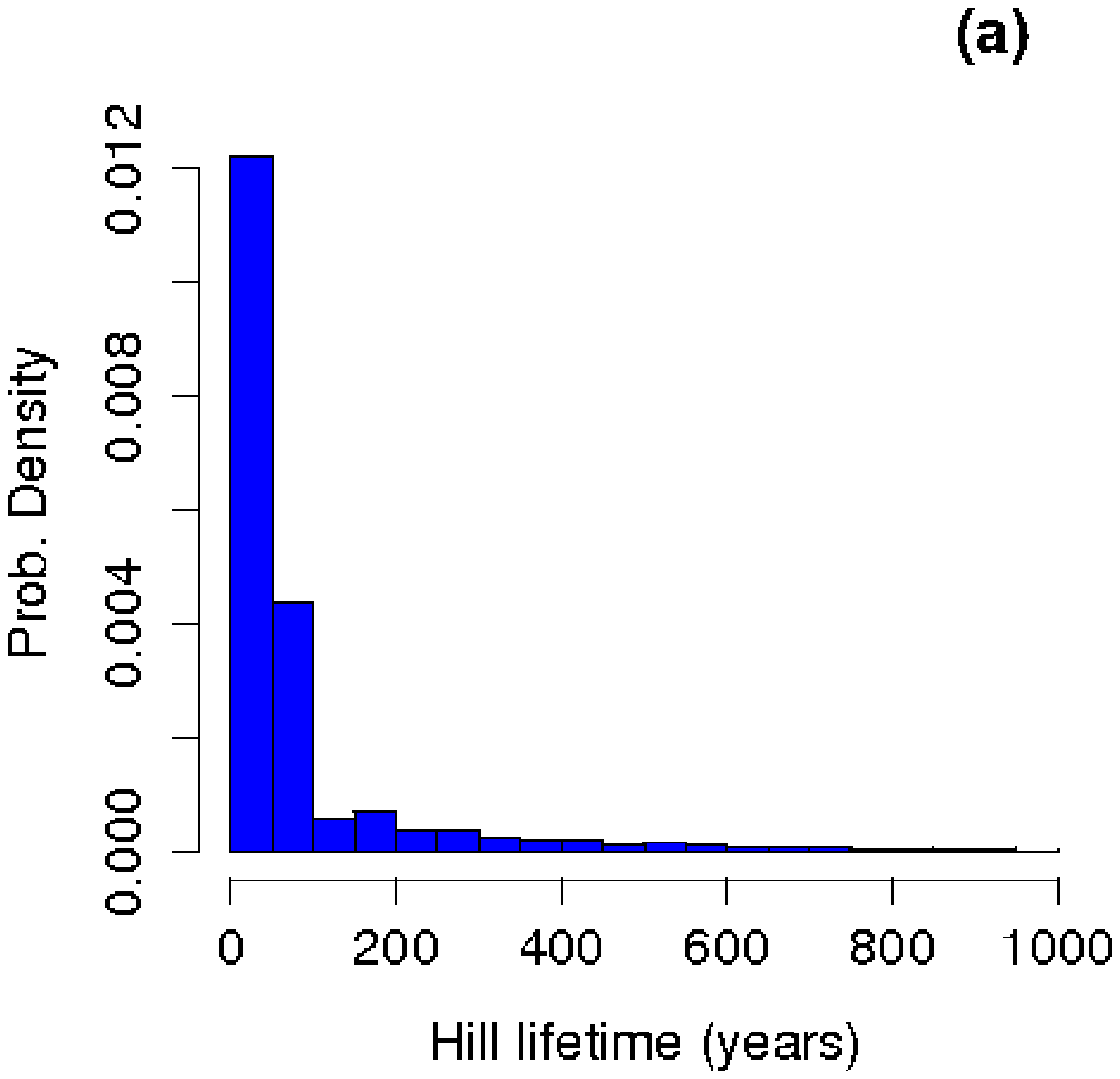}&
\includegraphics[scale=0.5]{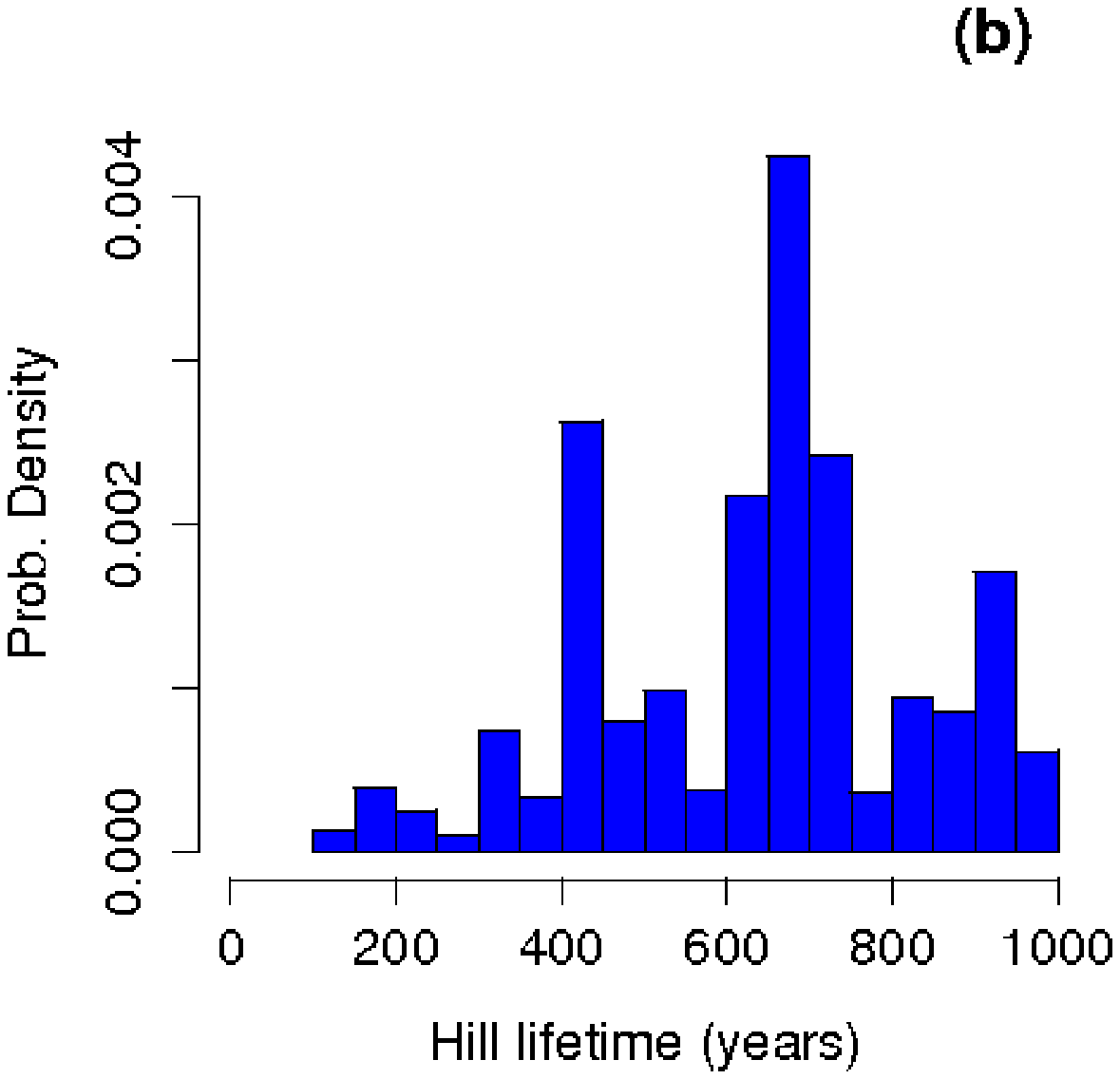}

\end{array}$
\end{center}

\caption{\label{pilot}Histograms of Hill lifetimes in the pilot calculations described in the text. Frame (a) shows the relative frequency of orbits with Hill lifetimes in the range $T_\text{Hill} \in{(0,1000)}$ years. Frame (b) shows the frequency of captured orbits as a function of the Hill lifetime. In computing the histogram in (b) equal numbers of initial conditions were used in each 100-year interval as described in the text.}
\end{figure*}

\item The mass ratio of the primaries is selected randomly and the stored orbital parameters of the primaries are rescaled using the actual values selected for $m_{1}$ and $m_{2}$. This is possible because, in the three-body Hill approximation, all masses can be scaled out of the problem \citep{Murray:1999}. However, the masses appear explicitly in the four-body Hill problem - see eq.\ (\ref{rho}).

\item The mass of the intruder is selected randomly up to the mass of the larger primary. Thus the maximum ratio of intruder mass to total binary mass is 0.5.

\item The intruder is then launched from infinity and the integration of the primaries simultaneously resumed - this time using the four-body equations of motion. The system is integrated until either the binary breaks up or it is stabilized. We have confirmed numerically that at the start of the four-body integrations the primaries and the intruder are sufficiently well separated that using re-scaled three-body Hill initial conditions to start the four-body integrations is legitimate. Thus, these computations cannot describe simultaneous ``three-body plus Sun'' encounters in which the primaries and the intruder all interpenetrate their mutual Hill sphere at about the same time. Of course, these encounters are describable by the four-body Hill equations but they are extremely rare and so are neglected.

\item Binary stabilization is registered if the binary survives for 10 times longer than its lifetime inside the Hill sphere in the absence of intruder scattering, or for $T = 200$ Hill units ($\sim$ 9640 years), whichever is larger. For example, if a transient had a Hill lifetime greater than 200 Hill units, say 201 Hill units or ~$\sim$ 9,693 years years then we counted it as stabilized only if it survived for at least 96,930 years. Capture statistics were quite insensitive to using lifetime extension multipliers larger than 10 but were somewhat sensitive to using multipliers $< 7$. As expected, intruders can prolong quasi-binary lifetimes by pushing trajectories deeper into chaotic zones but without actually stabilizing them permanently.

\item Keeping all masses fixed this procedure is then repeated for 5 different values of $R_0$.  This has the effect of sending the intruder towards the
binary at different relative configurations of the binary partners.

\item For each binary orbit the overall process was repeated for 1000 randomly selected intruders, each time varying all masses randomly.

\item  Individual integrations were stopped and results discarded if particles came within a distance $r_{A}=10^{-5}$ Hill units of each other. This radius is somewhat arbitrary but is roughly the radius of a typically-sized binary partner and thus this choice corresponds approximately to a physical collision between the binary members. Collisional singularities could be avoided by regularisation \citep{Szebehely:1967,Aarseth:2003} but we preferred to stop the integrations if a collision occurred. 

\item In total 15,000 quasi-bound binary orbits were harvested. For each binary, 1000 randomly selected intruders from a pool of 15,000 were then sent towards it as described above. Each binary-intruder encounter that led to capture was considered to be a single capture event.

\end{enumerate}

\subsubsection{Pilot calculations to identify a threshold Hill lifetime}

Examination of Fig.\ \ref{ltime}(a) reveals a potential computational difficulty associated with the algorithm just described. Ballistic trajectories - i.e., those which have Hill lifetimes $\sim 50 - 100$ years and follow hyperbolic or near-hyperbolic orbits - significantly outnumber trajectories which penetrate and become entangled in chaotic layers. Because, as we will show, these trajectories have very low capture probabilities, including them in the full simulations would swamp the calculations. On the other hand, we must first demonstrate that such capture probabilities are low. This is also important because the TNB formation mechanism of \citet{Goldreich:2002} assumes that it is precisely these trajectories which lead to capture, i.e., trajectories which transit the Hill sphere on time scales on the order of $R_H/V_H =1/\Omega_\odot \sim 50$ years.

We therefore performed a set of pilot calculations in which the capture of binaries with Hill lifetimes in a limited interval, i.e, $T_\text{Hill} \in{(0,1000)}$, years was studied. Figure \ref{pilot}(a) shows the initial distribution of Hill lifetimes obtained in this range using the procedure described above. It is apparent that if this distribution of orbits were to be used directly then the vast majority of the computations would involve trajectories with $T_\text{Hill} < 100$ years. In fact, simulations using this distribution had to be abandoned because most of the integrations of these initial conditions did not lead to capture. However, even in these limited simulations it was clear that the capture probabilities of trajectories with longer Hill lifetimes dominated despite their relative under representation in Fig.\ \ref{pilot}(a).

To study capture probabilities further we developed an alternative procedure.

\begin{enumerate}

\item 100 initial conditions were selected randomly in the ten intervals: $T_\text{Hill} \in{(0,100)}$ years, $T_\text{Hill} \in{(100,200)}$ years, etc., up to $T_\text{Hill} \in{(900,1000)}$.

\item 2000 intruders - selected as described above - were sent in towards these proto-binaries and the number of scattering events which led to capture in each interval was recorded.

\end{enumerate}

The results are shown in Fig.\ \ref{pilot}(b). The probability of capture for  $T_\text{Hill} < 100$ years is extremely small. Of course, in the original disk, primaries which enter their mutual Hill sphere and escape without being captured can, at a later time, re-enter the Hill sphere and  new opportunity to become caught-up in a chaotic layer presents itself. That is, escape is from the mutual Hill sphere and not from the disk itself. Therefore, even though the probability of a ballistic binary being captured in a single pass might be small repeated passes through the Hill sphere are possible. The overall capture probability will clearly depend on the mass and orbital element distributions in the original disk. This is not included in our simulations.

Based on Fig.\ \ref{pilot} we included only binaries with  $T_{\text{Hill}} \ge  200$ years in the large set of simulations now to be described. We note that the simulations in the pilot calculations and the full calculations show similar dependencies on intruder mass and binary mass ratio.

\section{Results}
\label{results}
\subsection{Effect of intruder mass}

\begin{figure*}
\begin{center}$
\begin{array}{cc}
\includegraphics[scale=0.3,angle=270]{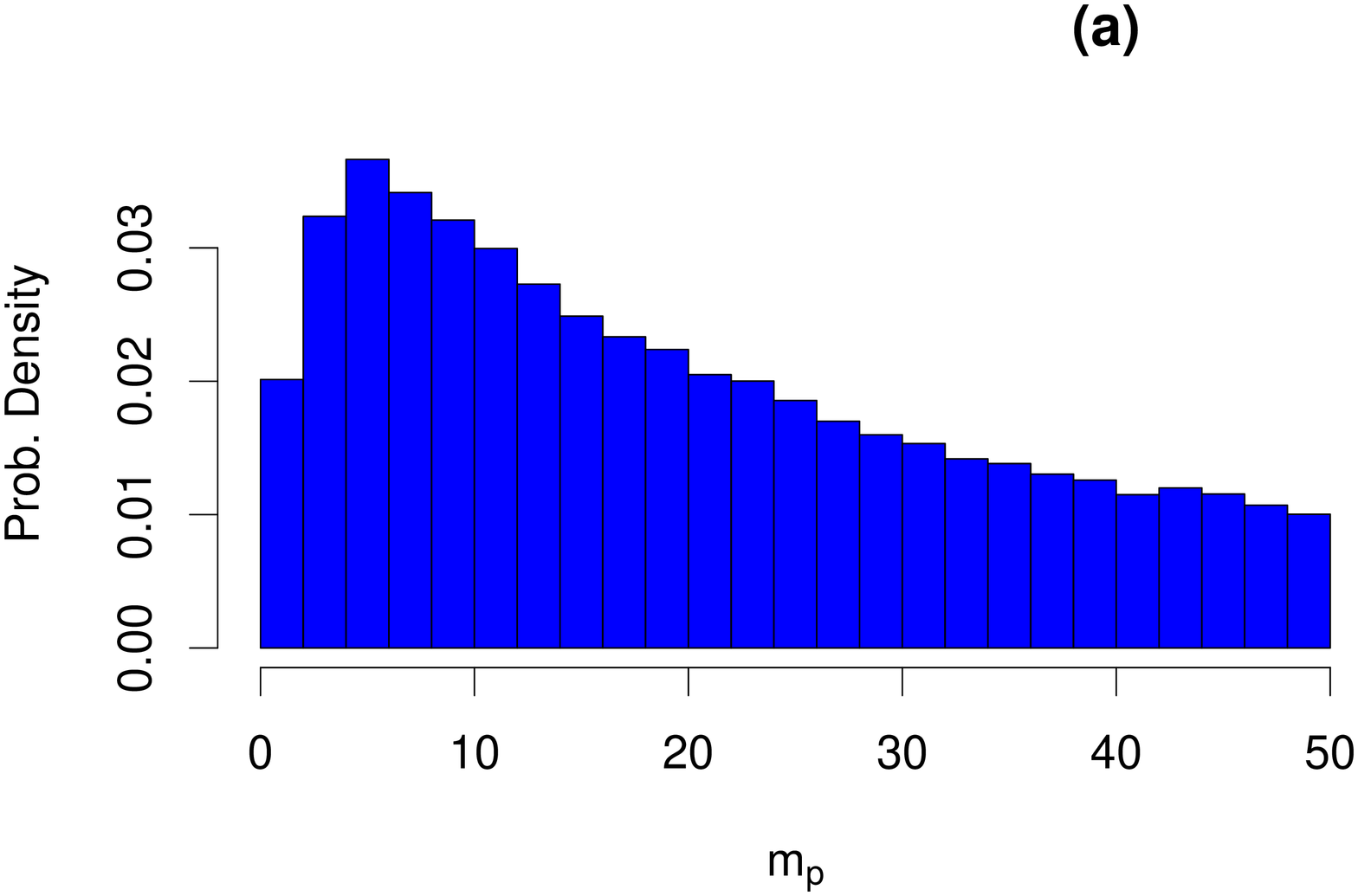}&
\includegraphics[scale=0.3,angle=270]{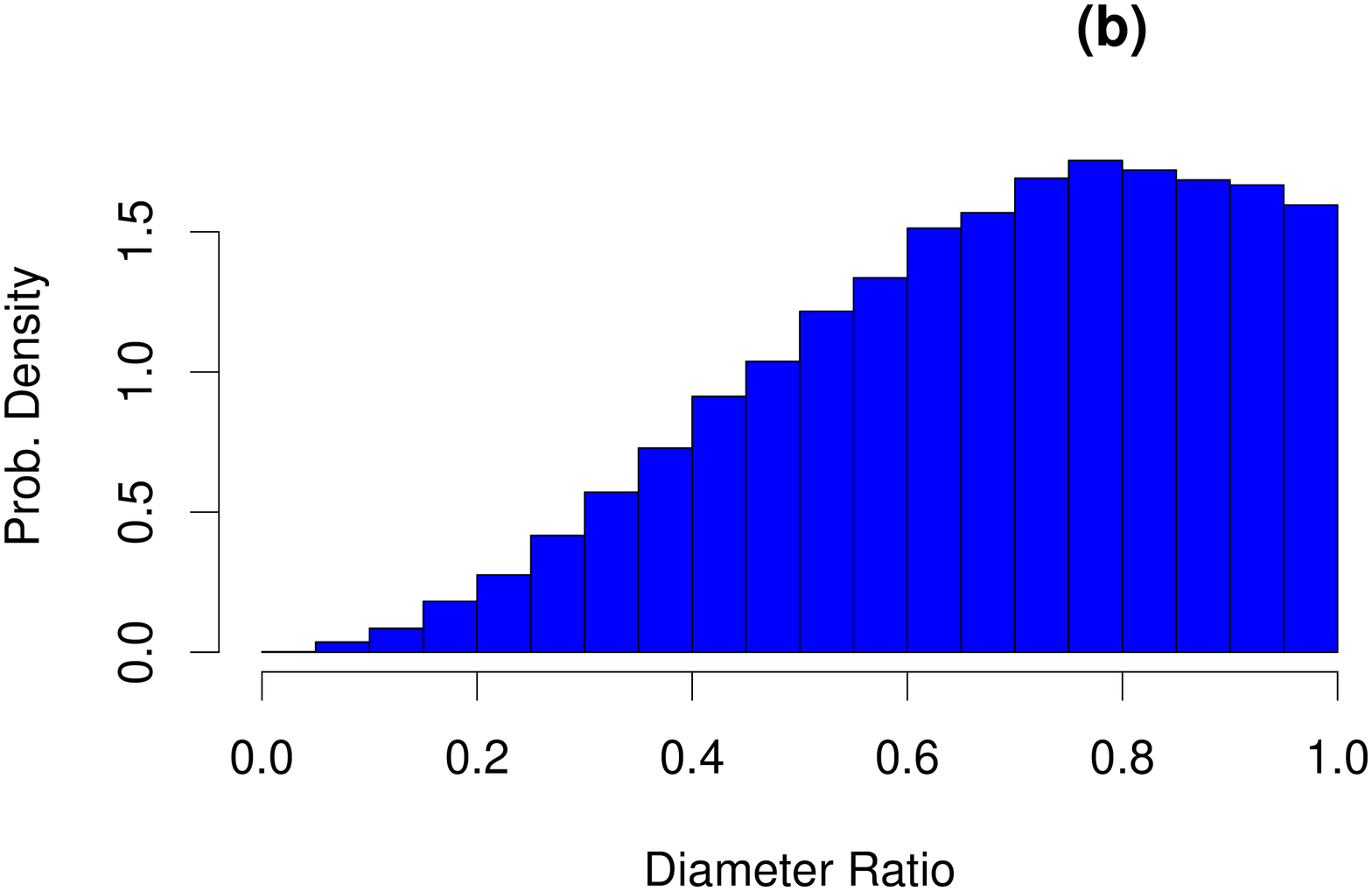}\\
\includegraphics[scale=0.3,angle=270]{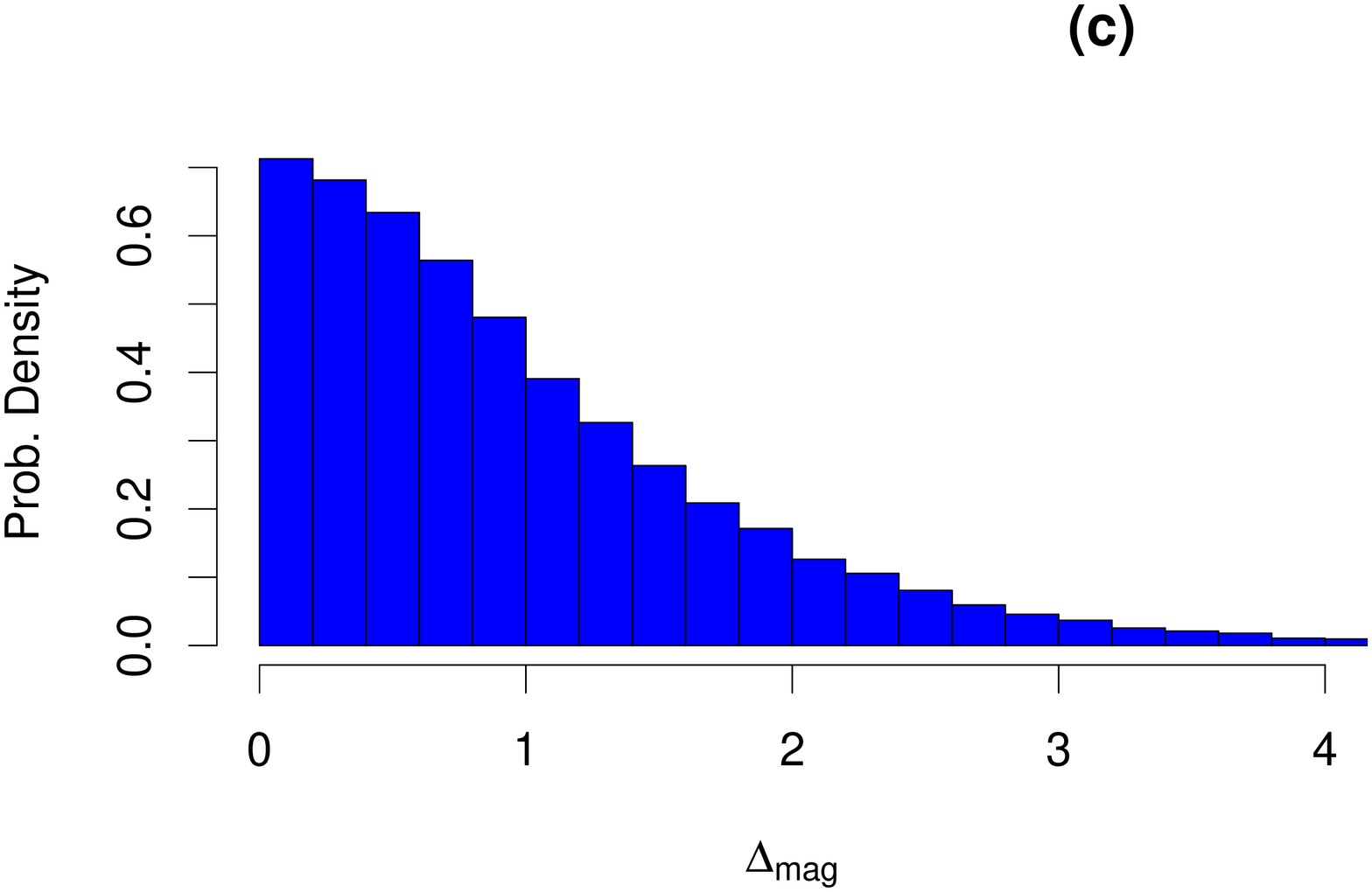}&
\includegraphics[scale=0.3,angle=270]{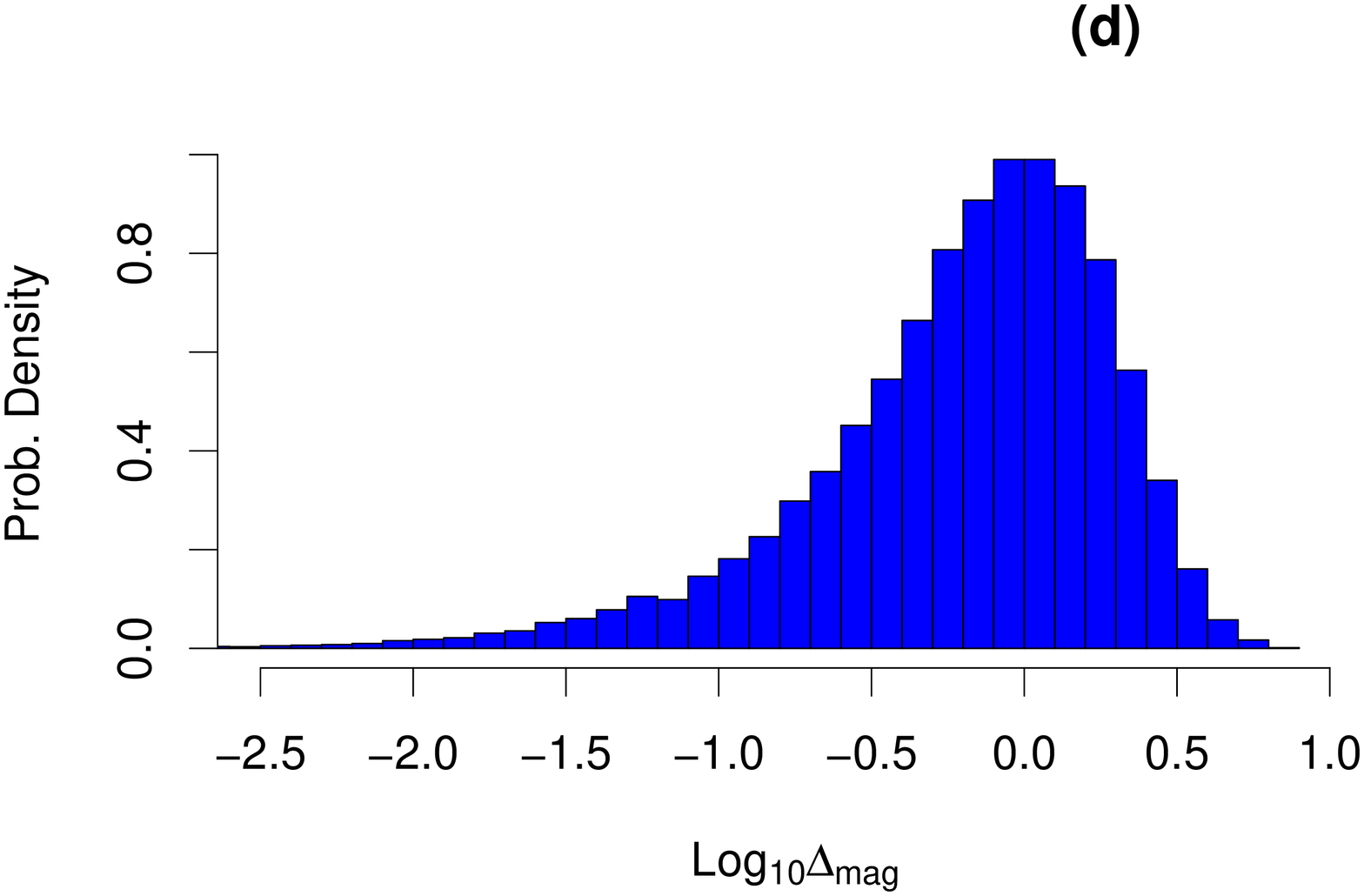}\\
\includegraphics[scale=0.3,angle=270]{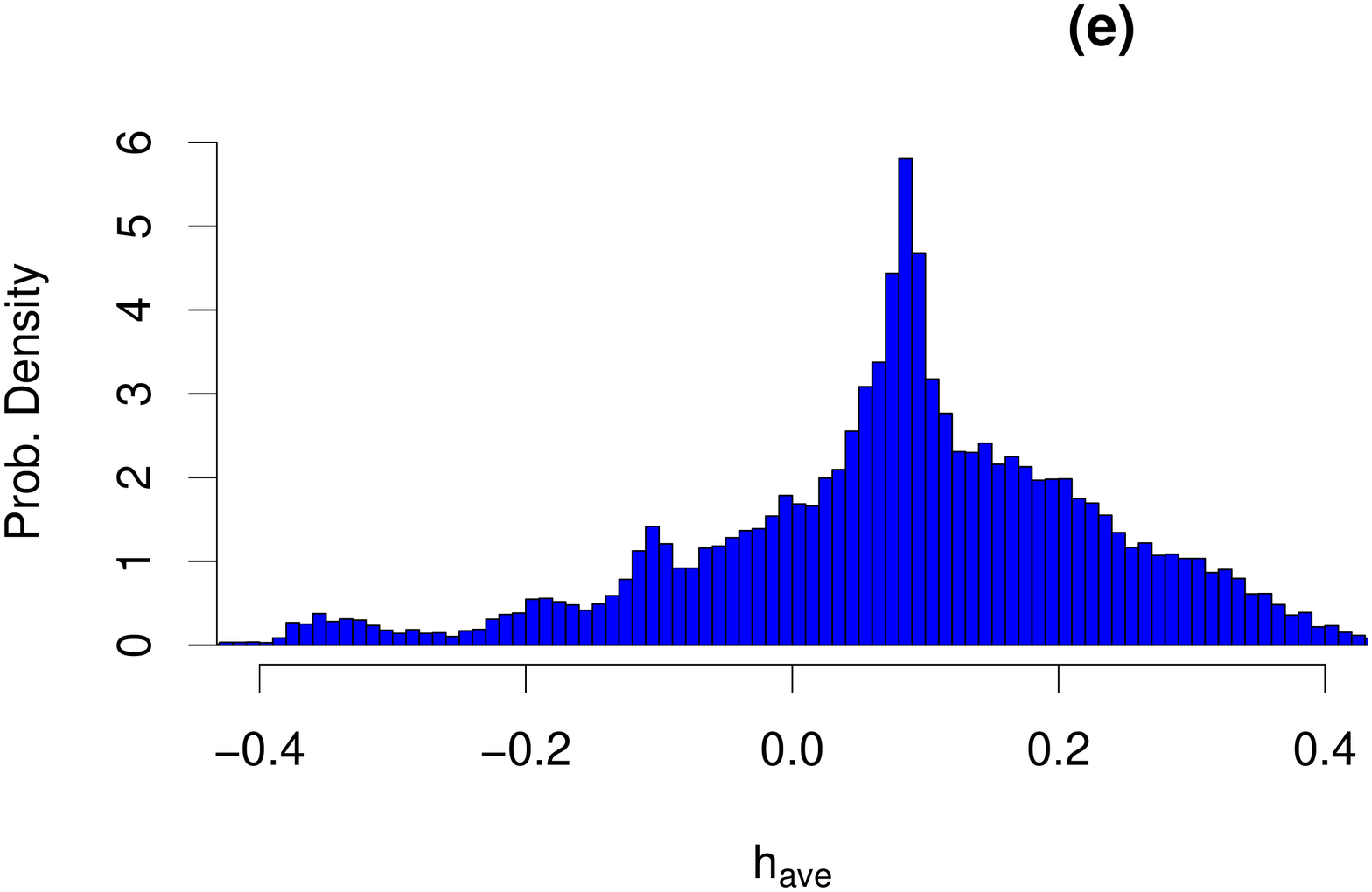}&
\includegraphics[scale=0.3,angle=270]{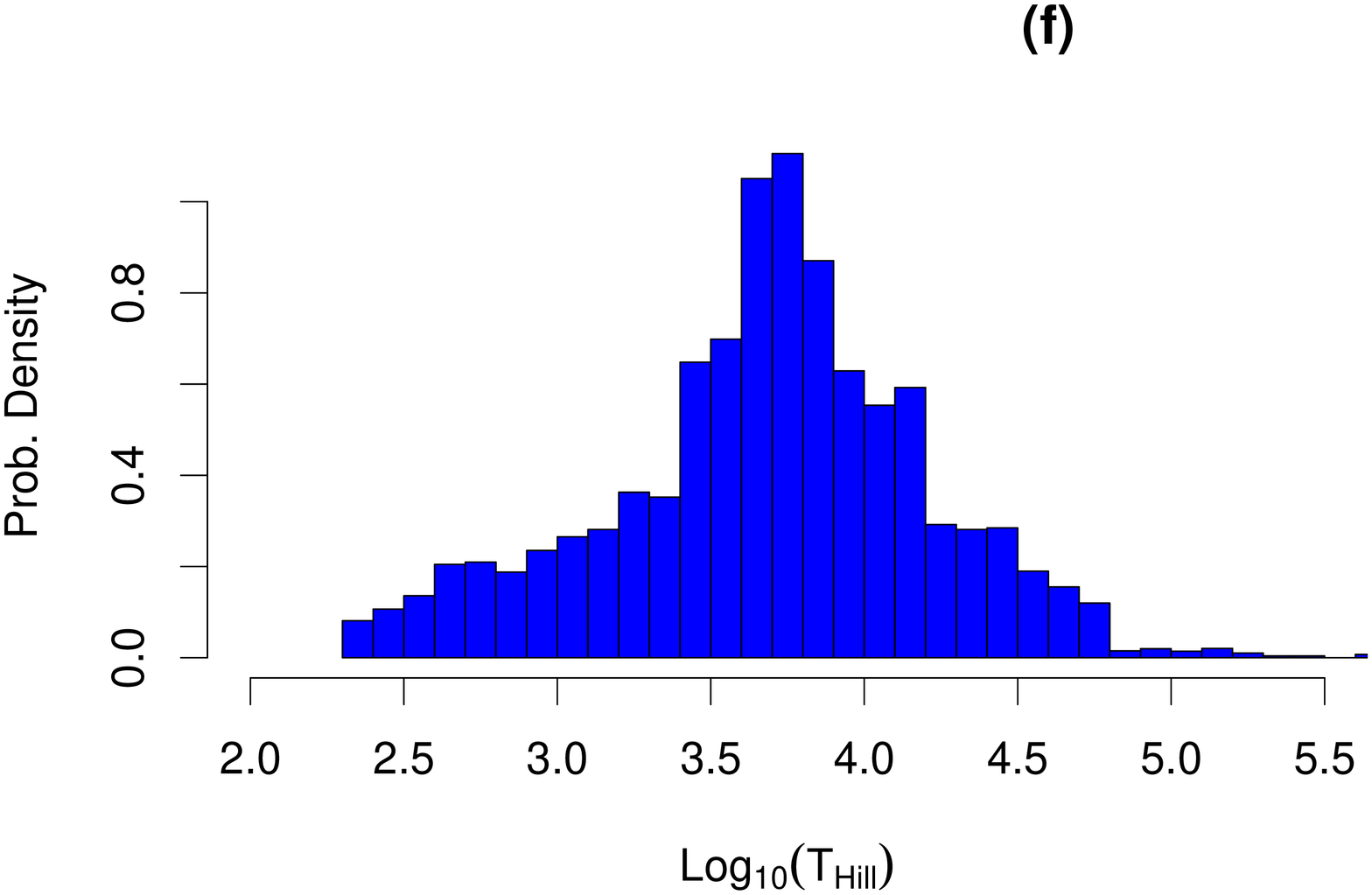}

\end{array}$
\end{center}

\caption{\label{4hists} Histograms showing the probability density - i.e., the area under each histogram is normalized to unity - of permanently captured binaries as a function of: (a) intruder mass where $m_p = m_3/(m_1+m_2) \times 100$ is the mass of the intruder expressed as a percentage of the total binary mass; (b) diameter ratio, $D_2/D_1$, of the primaries; (c) computed visual magnitude difference, $\Delta_{mag}$; (d) $\log_{10} \Delta_{mag}$; (e) average angular momentum, $h_{\text{ave}} = \langle h_\zeta \rangle$; prograde orbits correspond to $h_\zeta > 0$ and retrograde orbits to $h_\zeta< 0$, and; (f) $\log_{10}$ of their Hill lifetime in years, i.e., the maximum time the quasi-binary would spend inside the Hill sphere in the absence of intruder scattering.}
\end{figure*}

The masses of the primaries were varied randomly and the mass of the intruder was then chosen randomly up to the mass of the larger of the two primaries. Figure \ref{4hists}(a) indicates that the efficiency of binary capture falls off with increasing intruder mass, i.e., intruders of comparable size to the primaries tend to (i) leave the proto-binary essentially unaffected, (ii) destabilize the rather delicate quasi-bound binary (i.e., reduce its natural Hill lifetime by causing the complex to break up before its time) or (iii) stabilize it against ionization with lower probability than do smaller intruders. Destabilization happens, e.g., when a three-body resonance forms which has the effect of causing the binary to split up prematurely. However, we have also found some rarer cases where the intruder forms a resonance which actual lives longer than the Hill lifetime but which, nevertheless, eventually breaks up.

\begin{figure*}
\begin{center}
\includegraphics[scale=1,angle=0]{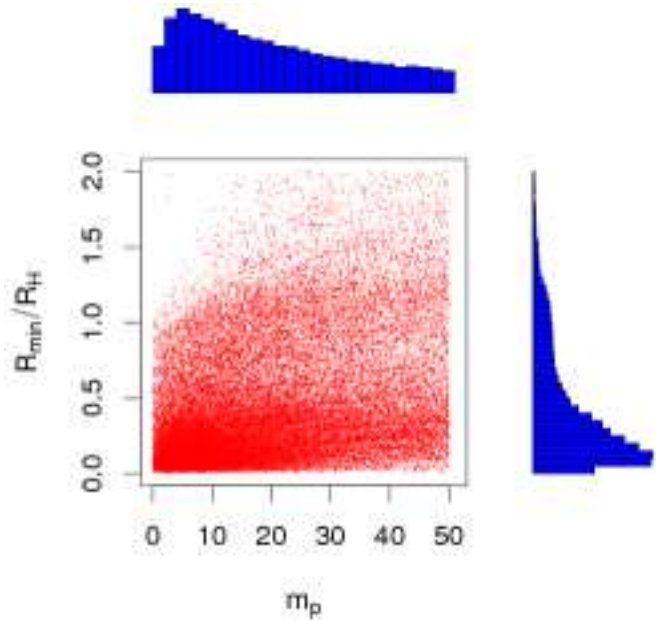}

\end{center}

\caption{\label{rmin} Scatter plot showing the minimum penetration distance, $R_{min}$,  of intruders which successfully capture binaries, as a function of intruder mass expressed as a percentage of the total binary mass. The attendant histograms show the distributions of $R_{min}$ and intruder mass. }
\end{figure*}

The rate at which the tail of the distribution (large intruders) falls off depends to some degree on the closest-approach distance of the intruder to the binary - see Fig.\ \ref{rmin}. That is, large intruders can be effective stabilizers provided that they do not get too close to the binary. On average, small intruders can penetrate much deeper and still stabilize the binary. It is apparent from Fig.\ \ref{4hists}(a) that small intruders are most efficient at stabilization. Given that roughly equal diameter binaries dominate in Fig.\ \ref{4hists}(b) we conclude that small intruders tend to lead to roughly equal mass ratios. Therefore, in view of Fig.\ \ref{rmin}, small $R_{min}$ values also correlate with roughly equal mass ratios. The reason for this effect is that small intruders can penetrate deeper before causing the binary to break up. For unequal mass binary partners this means that they can more effectively get caught-up in three-body resonances which tend to destabilize the binary.

\subsection{Effect of binary mass ratio}
As noted, the binary mass ratio was also varied randomly throughout the simulations. Assuming that the binary partners have the same density, mass ratios can then be translated into size- or diameter-ratios. Figure \ref{4hists}(b) shows a strong preference for roughly same-sized binary partners although this might be offset if a more realistic, e.g., a power-law, mass distribution were used.

\subsection{Magnitude differences}

Observed magnitude differences, $\Delta_{mag}$, of binaries can be used to obtain information about the relative sizes of the partners involved \citep{Trujillo:2001,Noll:2006a,Petit:2006}. Assuming that the primaries have the same density and albedo as each other \citep{Cruikshank:2005,Lykawa:2005,Stansberry:2006,Petit:2004} allows a relationship to be established between binary mass ratios obtained from calculations and observed $\Delta_{mag}$ values. The diameters of the primaries, $D_1, D_2$, are related to $\Delta_{mag}$ as follows \citep{Sheppard:2002,Hughes:2003}:

\begin{eqnarray}
\frac{D_2}{D_1} = 10^{-0.2\thinspace \Delta_{mag}}
\end{eqnarray}
where, by (our) definition, $D_2 < D_1$. Figures\ \ref{4hists}(c) and (d) show the predicted distributions of $\Delta_{mag}$, and (so as to amplify the region around $\Delta_{mag} = 0$) $\log_{10}(\Delta_{mag})$.

In order to make a more direct connection with actual observations \citep{Noll:2006a} it is important to know not only the magnitude differences but the magnitudes themselves. Our model makes no predictions about the absolute sizes of the binary partners - only their mass ratio is predicted. However, various fits have been made to the size distribution of TNOs with some suggestion that the number of objects with radii $\lesssim 40 - 70$ km \citep{Bernstein:2004,Bernstein:2006,Pan:2005,Elliot:2005,Petit:2006} is somewhat less than expected. \citet{Bernstein:2004} have proposed a double power-law fit to the differential surface density of TNOs as a function of magnitude. \citet{Petit:2006} have argued that a double power-law expansion is unnecessary although they provide parameters which fit their observations to the double power-law of \citet{Bernstein:2004}. For these reasons we used the single power-law distribution of \citet{Petit:2006} to simulate a plot of magnitude against magnitude differences. This was done as follows;
\begin{enumerate}
\item The differential magnitude distribution was multiplied by the efficiency function reported by \citet{Petit:2006}, which is a product of two hyperbolic tangents. This produced a ``corrected'' distribution \citep{Trujillo:2001,Elliot:2005}.
\item This distribution was normalized according to the number of captured orbits obtained from the simulations ($\sim 70, 000$). 
\item Using this distribution visual magnitudes were randomly assigned to the larger member of each nascent binary with magnitudes, $m$, in the range $20 \le m \le 30$.
\item A bootstrapping procedure was used in which the previous step was repeated 5 times using a different random number seed in each case. Thus, each captured binary actually enters in with 5 different randomly assigned magnitudes. This procedure improves the statistics for smaller magnitudes where comparatively few instances are generated.

\end{enumerate} 

\begin{figure*}
\begin{center}
\includegraphics[scale=0.25]{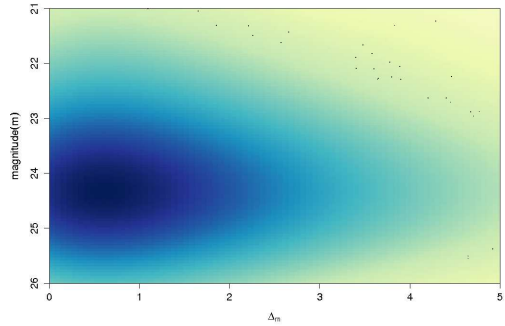}

\end{center}

\caption{\label{dmag} (colour online) Kernel smoothed colour density scatterplot showing the visual magnitude (mag.) of the larger binary primary as a function of magnitude difference, $\Delta_{\text{mag}}$.  The distribution used to assign visual magnitudes is modified by a double hyperbolic tangent efficiency function \citep{Petit:2006}. A total of  $\approx$ 70,000 captured binary orbits were used over a magnitude range $20 \le m \le 30$. Colour scale runs from low [yellow (online), light grey (print)] to high [dark blue (online), dark grey (print)] and the kernel smoothing is as described in Fig.\ \ref{ltime} and Appendix A.}
\end{figure*}

The results are shown in Fig.\ \ref{dmag}. Keeping in mind the caveats already noted in regard to albedo and density, the results in Fig.\ \ref{4hists} and Fig.\ \ref{dmag} demonstrate that the capture model predicts a similar clustering of $\Delta_{mag} <  1$ to that observed by \cite{Noll:2006b,Noll:2007}. In our model, the physical reason is chaotic destabilization of unequal mass quasi-binaries in comparison to those having roughly equal masses \citep{Astakhov:2005}.

\subsection{Binary orbit mutual inclinations}

Strictly speaking it is not possible to compute meaningful orbital elements for nascent, i.e., just-formed, binaries. This is because these objects generally follow highly non-Keplerian orbits \citep{Astakhov:2005} and their orbital elements will change due to post-formation hardening. However, in previous studies of the CAC model we have found that the sign of the angular momentum component, $h_\zeta = \xi \thinspace p_\eta - \eta \thinspace p_\xi$, tends to be roughly conserved and that the inclination is quite robust to post-capture hardening and orbit reduction though gas-drag \citep{Astakhov:2003} or gravitational scattering \citep{Astakhov:2005}. However, the inclination distribution may be sensitive to the hardening mechanism as well as other assumptions in the model. Further investigation is called for.

During the transition from proto-binary (slightly unbound) to nascent binary (slightly bound) the inclination hardly varies during the rather ``delicate" interaction with an intruder (this is not always the case, but it is the case more often than not). Thus the inclination distributions before and after capture are quite similar. This is a direct consequence and signature of CAC because it is the location of the chaotic layers in phase space that determines which inclinations are most favourable for a given value of the Jacobi constant \citep{Astakhov:2003,Astakhov:2004,Astakhov:2005}.

For each nascent binary we computed its average post-capture angular momentum, $h_{\text{ave}}=\langle h_\zeta \rangle$. These results are shown in Fig.\ \ref{4hists}(e). The distributions are quite asymmetric, especially in the tails, and predict a broad distribution of orbit inclinations. Although difficult to determine observationally, this suggest that mutual inclination distributions will provide a strong test of our model. As noted, this result is similar to the CAC model of irregular satellite capture which predicts different sub-populations of retrograde and prograde moons at Jupiter and Saturn \citep{Astakhov:2003,Hamilton:2003}. However, further study of the details of the effect of hardening on inclination distributions is clearly merited.

\section{Comparison with the L$^2${\small{s}} and L$^3$ models}
\label{discussion}

 \citet{Goldreich:2002} estimated binary formation rates corresponding to two different mechanisms. The first mechanism, denoted ``L$^2$s,'' involves dynamical friction where ``L'' refers to a ``big'' body and ``s'' to a sea of small bodies which contain most of the mass. Essentially, two big bodies (L $+$ L) enter their mutual Hill sphere and become permanently captured by the dynamical friction exerted by the sea of small bodies. The second mechanism, denoted ``L$^3$,'' involves gravitational scattering by a third big body, i.e., 
\begin{eqnarray}
\text{L}^{2*} + \text{L} \rightarrow \text{L}^2 + \text{L}^* 
\end{eqnarray}
which extracts energy ($^*$) from the proto-binary thereby stabilizing it.

\subsection{The L$^2$s mechanism}

Dynamical friction, in principle, provides
an efficient route to binary hardening and orbit reduction. However, it is not sensitive to the mass ratio of the
initial quasi-bound binary. The Hill equations,
even including dissipation, can be scaled so as to contain no parameters \citep{Astakhov:2003,Murray:1999} and so do not
depend on the masses of the binary partners. Thus, no mass selection
is predicted.  In addition, estimates of planetesimal mass distributions suggest that L$^2$s might not be an efficient binary production mechanism \citep{Funato:2004}.

\subsection{The $L^{3}$ mechanism}
In $L^{3}$, as originally proposed, roughly
equal mass binaries result directly from the assumption that a
transient binary undergoes a scattering encounter with a third
body of comparable mass to the binary partners (which are also assumed
to be of similar mass). In essence, the preference for
order unity mass ratios is partly a result of the
assumptions of the model. Furthermore, multiple $L^{3}$
encounters are assumed to harden the binary. We have shown previously
that multiple large body encounters tend to
disrupt the binary \citep{Astakhov:2005}. Examination of the
actual orbital elements produced in these simulations indicate
that neither a single nor a small number of $L^{3}$-type events is
sufficient to produce Keplerian orbits. Those simulations - together with the current set - suggest that the assumption that intruder scattering stabilizes binaries with a probability of order
unity is a considerable overestimate, especially for binaries which pass through the Hill sphere 
on time scales of order $R_H/V_H$.

However, both of the mechanisms presented by \citet{Goldreich:2002} yield binary formation rates on the order of 10$^{-6}$ or 10$^{-7}$ per year which are comparable to large body growth time scales and are, therefore, quite reasonable in themselves. However, these numbers depend sensitively on assumptions made about various quantities, e.g., the residence time of a transient binary inside the Hill sphere and, in the case of L$^3$, the probability of a fourth body entering the Hill sphere and stabilizing the transient binary during its Hill lifetime. Getting numerically reasonable rates does not necessarily imply that the underlying mechanism and assumptions are reasonable. In fact, the CAC model helps to obtain better estimates for some of these quantities and these estimates can then be used to generate better rate estimates. Using these estimates, CAC seems to produce KBB formation time scales which are not unrealistically long and, using our estimated parameters, are, in fact, on the order of 10$^{-6}$ or 10$^{-7}$ per year or better.

Another problem has been identified by \citet{Chiang:2006} and \citet{Noll:2007} with the L$^2$s and L$^3$ mechanisms; assuming that the encounter time of the primaries in the Hill sphere $t_{\text{enc}} \sim R_H/V$ then, for velocity dispersions with $v > v_H$ the time scale for binary formation goes as $(v/v_H)^{11}$. On the other hand, if velocity dispersions are on the order of $v_H/3$, i.e., $v < v_H$, as assumed by \citet{Goldreich:2002}, then big bodies should collapse into a vertically think disk \citep{Goldreich:2004} with mutual orbit normals parallel. This would imply that binary inclinations should be similar when, in fact they seem to be randomly distributed.

In this section we compare CAC formation rates using similar ``back-of-the-envelope'' arguments to those provided by \citet{Goldreich:2002} and examine the fine-tuning problem. It is important to stress, however, that this discussion is not an attempt to compare the two models based on raw predictions of formation rates. The main aim is to determine if CAC leads to rates similar to those obtained by \citet{Goldreich:2002} which, numerically, seem reasonable. In particular, based on our simulations we conclude that several assumptions in the model of \citet{Goldreich:2002} may lead to overestimates of binary formation rates. 
\subsection{Binary formation rates}
Formation rates can be estimated in the L$^2${s} and L$^3$ models by making a few apparently reasonable assumptions. Both processes (and also the CAC mechanism) involve the rate at which two big bodies enter their mutual Hill sphere. Assuming that the debris disk is infinitely thin, then the entry rate per big body can be estimated as being the product of three factors \citep{Goldreich:2002,Goldreich:2004}:
\begin{eqnarray}
\dot{\Theta} \approx \left( \frac{\Sigma}{M} \right) \times v_{big} \times R_H \label{hillent}
\end{eqnarray}
where $\Sigma$ is the surface mass density of big bodies of mass $M$. The first term is the number of big bodies per unit area. The second term is the velocity while the third is an estimate of the range of impact parameters over which entry into the Hill sphere occurs. 

If it is assumed that bodies approach the Hill sphere with approximately the Hill velocity then eq. (\ref{hillent}) becomes
\begin{eqnarray}
\dot{\Theta} \approx \left( \frac{\Sigma}{M}\right) \thinspace R_H^2 \thinspace \Omega_\odot \approx \left(\frac{\Sigma \thinspace \Omega_\odot} {d \thinspace R}   \right)
\times \frac{1}{\theta_\odot^2} \label {hillent1}
\end{eqnarray}
where eq. (\ref{hillv}) has been used. Here, $d$ is the big body density and $\theta_\odot$ is essentially a scaling parameter which relates the radius of a big body to the Hill sphere radius \citep{Goldreich:2004}; i.e., $\theta_\odot = R_{\text{big}}/R_H \sim 10^{-4}$; it turns out to be the angular size of the Sun as seen from the Kuiper belt \citep{Goldreich:2004}
The overall formation rate is the product of the Hill entry rate ($\dot{\Theta}$) and the probability ($P$) that an encounter with an intruder leads to capture. Assuming prompt or ballistic transport through the Hill sphere leads to an average transient binary lifetime of $t_{life} \approx R_H/V_H = 1/\Omega_\odot$. The probability that a third large body will enter the quasi-bound binary Hill sphere during its lifetime is, therefore,
\begin{eqnarray}
P \approx \left(\frac{\Sigma \thinspace \Omega_\odot} { d \thinspace R_{\text{big}}}   \right)
\times \frac{1}{\theta_\odot^2} \times \frac{1}{\Omega_\odot}.
\label{cap1}
\end{eqnarray}
The time scale for binary formation based on these two-body types of estimates becomes 
\begin{eqnarray}
T_{\text{2B}} \approx \frac{1}{p\thinspace P \thinspace \dot{\Theta}} \approx \frac{1}{p}\left( \frac{ d \thinspace R_{\text{big}}}{\Sigma} \right)^2 \thinspace \frac{\theta_\odot^4}{ \Omega_\odot}
\label{timescale}
\end{eqnarray}
where $p$ is the probability that, having entered the Hill sphere, a capture will occur. Setting $p=1$ and using the parameters suggested by \citet{Goldreich:2002} or \citet{Chiang:2006} leads to a time scale on the order of 2 - 3 Myr for a quasi-bound binary to fuse.

There are several difficulties with the approach just described, particularly in regard to computing the binary lifetime inside the Hill sphere. Binaries which transit the Hill sphere promptly, i.e., on a time scale $\approx R_H/V_H \sim 50$ years are rarely captured. Figure \ref{ltime} shows that different trajectories having the same asymptotic velocity can spend very different times inside the Hill sphere depending on the particulars of their orbital elements. Further, small asymptotic velocities need not imply slow transit of the Hill sphere (see Fig.\ \ref{ltime} - the actual velocity after entering the Hill sphere can be substantially higher than the asymptotic value. While ``rapid transit'' binaries do indeed contribute to the overall entry rate into the Hill sphere their lifetimes are not relevant for computing the encounter rate of quasi-bound binaries with intruders. This is because chaos can induce a substantial time delay in the transit time of a fraction of the entering binaries. The lifetimes of these ``resonant'' states can be several orders of magnitude larger than is implied by prompt transport as Fig.\ \ref{ltime} shows. Because their formation is a non-linear process these lifetimes cannot be predicted in any simple way. The situation is somewhat similar to a chemical reaction in which resonances - essentially the time-delay due to the formation and eventual  decay of a quasi-bound state - can have a profound effect on the reaction cross-section \citep{Levine:1987,Althorpe:2004}.

The possibility of very long-living resonant quasi-bound binaries suggests that an alternative - though still ``back-of-the-envelope'' style \citep{Goldreich:2002,Chiang:2006} - treatment of binary formation rates might be useful.

Assuming that the overall entry rate is roughly given by eq.\ (\ref{hillent1}) we need to estimate the fraction of entrants which get caught up in chaotic layers as well as their lifetimes therein - i.e., their Hill lifetime. Suppose that the probability of a trajectory having a lifetime equal to, or greater than, some threshold Hill lifetime $T_{0}$ is $P_{>}$. Then the combined probability that a third large body will enter the quasi-bound binary Hill sphere and capture the binary permanently during its lifetime $T > T_{0}$ becomes
\begin{eqnarray}
P \approx  \left(\frac{\Sigma \thinspace \Omega_\odot} { d \thinspace R_I}   \right)
\times \frac{1}{ \Omega_\odot \thinspace \theta_\odot^2} \times P_{>} \times  T_{0} \times p \times v_I \label{cap2}
\end{eqnarray}
where $R_I$ is the radius of the intruder and $v_I$ is the intruder velocity in units of $v_H$; we take $v_I \sim 2 v_H$ - see Fig.\ \ref{ltime}. This is the estimate of \citet{Goldreich:2002} multiplied by a correction factor due to the delay inside the Hill sphere due to CAC. Although, in fact, $p < 1$, in order to make a direct comparison with estimates made by \citet{Goldreich:2002} and \citet{Chiang:2006} we also set $p = 1$ and $R_I = R_{\text{big}}$. This likely underestimates the actual capture time scales in the two-body simulations.

\begin{figure*}
\begin{center}
\includegraphics[scale=0.25,angle=270]{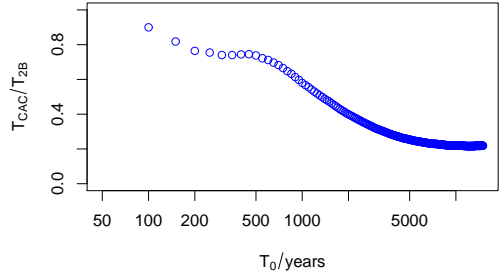}

\end{center}

\caption{\label{thill} Ratio of two different estimates of the time scale for binary formation: $T_{\text{CAC}}$ is the CAC time scale and is estimated using eq.\ (\ref{cap2}); $T_{\text{2B}} \sim 2- 3$ Myr is the corresponding two-body estimate from eq.\ (\ref{timescale})  \citep{Goldreich:2002,Chiang:2006}; $T_0$ is the threshold Hill lifetime (logarithmic scale) - see discussion surrounding eq. (\ref{cap2}).}
\end{figure*}

The cumulative probability $P_{>}$ can be calculated directly from our simulations and thus the increase or decrease in the timescale for binary formation as compared to eq.\ (\ref{timescale}) can be estimated as a function of $T_{0}$. Figure \ \ref{thill} shows the ratio of the thusly computed CAC time scale ($T_{\text{CAC}}$) and the corresponding two-body estimate based on eq.\ (\ref{timescale}), i.e., $T_{\text{2B}}$. We conclude that the time scales for binary formation in the CAC model are on the order of a few Myr. 

So far the discussion has neglected the small-intruder effect apparent in Fig.\ \ref{rmin}. If we assume that the intruder has a mass $\sim 2 \%$ of the total binary mass then $R_I \sim R/3$ and, for long-lived transient binaries we estimate from the simulations that $p \sim 0.3$ which, together, decreases the time scale for capture by a further factor of $\sim 3$.  Of course, decreasing $p$ in eq. (18) seems to increase the CAC time scale compared to estimates by \citet{Goldreich:2002}. However, it should be noted that in their work $p$ was set to unity - i.e., the assumption was made that {\it all} intruder scattering events would lead to capture. Our results indicate that $p < 1$ and, for binaries which transit the Hill sphere rapidly - i.e., those which do not become entangled in chaotic layers - $p <<1$. Thus, in principle $p$ should also be reduced when computing the two-body rates.

Nevertheless, the CAC model, even though it permits supra-$v_H$ velocity dispersions, and relies on the formation of relatively rare long-lived transient binaries, leads to capture rates which seem reasonable, i.e., on the same order-of-magnitude time scale as the growth of isolated big bodies \citep{Goldreich:2002,Rafikov:2003c,Goldreich:2004}.

\subsubsection{Role of the Hill velocity}

Several observations on the significance of the Hill velocity in these simulations (and rate estimates) can be made. 

\begin{enumerate}

\item The Hill velocity is defined to be the orbital velocity around a large central object at the Hill sphere radius.  The assumption in analytical or semi-analytical estimates of binary formation rates \citep{Goldreich:2002} is that if binary components are separated by $\approx R_H$ then they orbit each other with the same period as the binary center of mass orbits the Sun, i.e., $\approx 1/\Omega_\odot$ \citep{Goldreich:2004,Chiang:2006}. However, this will rarely be the case, especially if the binary partners are following a chaotic orbit: galleries of orbits which inhabit outer regions of the Hill sphere \citep{Murison:1988} together with their surfaces of section \citep{Winter:1994,Winter:1994a} are available and these indicate a broad range of possible orbits, few of which seem likely to satisfy the Hill velocity criterion. In fact, the stability radius for two-body mutual motion inside the Hill sphere is considerably less than $R_H$ \citep{Innanen:1979,Toth:1999,Nicholson:2006,Szebehely:1967}; although stable orbits can be found at very large fractions of $R_H$  their motion is highly non-Keplerian. Therefore, it is unclear to what extent the Hill velocity - an essentially two-body quantity - is useful in this situation. 

\item In view of the previous observation, and from the simulations, even if the velocity at infinity is $\sim v_H$ then this does not imply that the entry speed of the primaries - or the intruder - into the Hill sphere will be comparable to $v_H$. In addition, the instantaneous relative velocity of the quasi-bound primaries can be significantly larger than $v_H$.

\item After the particles enter the Hill sphere what happens next is inherently unpredictable except by explicit computation. Primaries which become entangled in chaotic layers may have very long lifetimes whereas others may transit promptly and it is difficult, in a chaotic system, to distinguish the two classes {\it a priori}. It seems likely that more sophisticated phase space transport theories will be needed to estimate rates semi-analytically \citep{Jaffe:2000,Jaffe:2000a,Jaffe:2002,Uzer:2002,Waalkens:2005}. In principle, these rates will be related to the relative volumes of chaotic and scattering parts of phase space \citep{Jaffe:2000a}. 

\subsection{The fine-tuning problem}

\citet{Goldreich:2002} have estimated that the velocities of $\sim 100$ km sized bodies were on the order of $v_H/3$ in the early Kuiper-belt. However, \citet{Chiang:2006} and \citet{Noll:2007} note that if the velocity dispersion, $v$, of big bodies is less than $v_H$ then these bodies will collapse into an essentially two-dimensional disk due to runaway cooling which is caused by the dynamical friction exerted by the sea of planetesimals. If the primordial binary partners inhabited a vertically thin disk then TNB mutual orbit inclinations ought to be similar, i.e., mutual orbit normals should be approximately parallel. In fact, TNB inclinations appear to be randomly distributed \citep{Noll:2003,Chiang:2006} which implies an isotropic primordial velocity distribution.  This suggests that the big bodies did not all originally lie in the same plane \citep{Noll:2007} and, therefore, that binaries formed when $v > v_H$. However, assuming that $v > v_H$ has an unexpected snag.

\citet{Noll:2007} have demonstrated that rate estimates based on two-body dynamics predict an extremely sharp decrease in capture probability for velocity dispersions which exceed $v_H$ even slightly \citep{Noll:2007} - according to their work, capture time scales go as $(v/v_H)^{11}$ which essentially precludes binary formation unless $v \le v_H$.  \citet{Noll:2007} obtain this result by making the following assumptions:
\begin{enumerate}
\item For $v > v_H$ the actual ``sphere-of-influence'' of the primaries is not $R_H$ but $R_I$ where

\begin{eqnarray}
R_I =  R_H \times (\frac{v_H}{v})^2
\end{eqnarray}
\item For $v > v_H$ the average encounter time (in years) inside the sphere-of-influence is $t_\text{enc} \sim R_I/V$

\end{enumerate}

As velocities increase the sphere-of-influence shrinks as does $t_\text{enc}$; powers of $v/v_H$ proliferate and capture time scales $\sim (v/v_H)^{11}$ result.  Obviously, according to this estimate, even a small increase of $v$ above $v_H$ will lead to dramatically long time scales for binary formation. This constitutes the fine-tuning problem \citep{Noll:2007}; observed inclinations suggest that $v\gtrsim v_H$ but rate computations indicate that $v \le v_H$. Together, this implies that $v = v_H$ which is an example of fine-tuning. As \citet{Noll:2007} ask ``Why should $v \approx v_H$ during binary formation?" There seems to be no good reason.

The CAC model does not suffer from a fine-tuning problem which, at its heart, springs from the assumption that the encounter lifetime goes as $R_H/V$. This is for two reasons:
\begin{enumerate}
\item Figure \ref{rmin} shows that, as predicted by \citet{Noll:2007}, the ``sphere of influence" is reduced in size as compared to the Hill sphere radius. Intruders which penetrate deeper have a higher probability of leading to capture.  Nevertheless, the steepest drop-off in density in Fig.\ \ref{rmin} occurs just outside $R_H$ which suggests that setting $R_I = R_H \times (v_H/v )^2$ (when $v > v_H$) \citep{Noll:2007} is probably too severe of an approximation. Figure \ref{ivel} shows the distribution of intruder asymptotic velocities leading to capture. These velocities, typically, are comparable to or exceed $v_H$. However, small intruders in the primordial Belt were likely following more elliptical orbits and therefore their velocities may have substantially exceeded $v_H$. Simulations in the elliptical Hill four-body problem will be needed to model such encounters.
\begin{figure*}
\begin{center}$
\begin{array}{c}
\includegraphics[scale=0.3,angle=270]{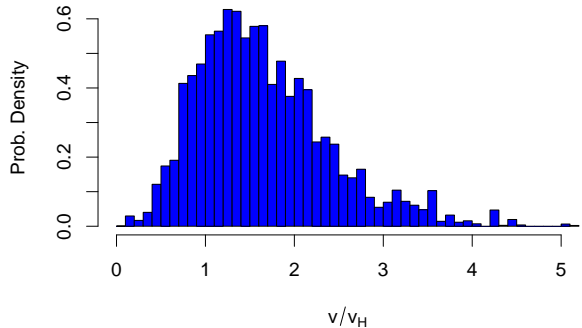}
\end{array}$
\end{center}

\caption{\label{ivel} Intruder velocity distribution leading to capture. Units scaled by the Hill velocity $v_H$.}
\end{figure*}
\item In CAC the encounter lifetime, $t_\text{enc}$ need not bear much relation to $v$ - see Fig.\ \ref{ltime} - and, in fact, the encounter lifetime inside the Hill sphere can be significantly extended by capture into chaotic layers. Thus, CAC is consistent with randomly distributed binary inclinations because it allows significant capture probabilities when $v > v_H$, i.e., it does not suffer from a fine-tuning problem. 

\end{enumerate}

From these considerations we conclude that the CAC mechanism is, in principle, consistent with observed TNB mutual orbit inclination distributions. Because the Hill lifetime can be quite different from $R_H/v$ it seems likely that the origin of the fine-tuning problem lies in the essentially two-body estimates which constitute its basis - it appears that the fine-tuning problem was raised by \citet{Noll:2007} precisely to make this point.
\end{enumerate}

\subsection{Mass selection: Comparison with L$^2$s and L$^3$ }
The prediction of a preference for similarly sized binary partners in the CAC model departs from both the L$^2$s dynamical friction formation model and the L$^3$ model \citep{Goldreich:2002}. 

L$^2$s: For the small (relative to the Sun) masses of the TNB partners the Hill approximation is extremely accurate. In fact, as noted, the individual masses of the two small bodies - and the total mass - can be scaled out of the equations of motion entirely \citep{Petit:1986,Henon:1986,Murray:1999}. Therefore, simple dynamical friction cannot select for any particular mass ratio; in order for dynamical friction to be effective it suffices that the {\it total} mass of the potential binary be much larger than the individual masses of planetesimals which make up the sea of small bodies exerting dynamical friction. This can be seen by noting that the semi-major axis which undergoes steady reduction through dynamical friction in the model of \citet{Goldreich:2002} is actually the semi-major axis of the mutual orbit in the Hill approximation. Thus, while the L$^2$s model of \citet{Goldreich:2002} seems to require large primaries, it is only necessary that the {\it total} binary mass is large as compared to planetesimal masses. Therefore, this model does not make predictions about the mass ratio of the binary partners.

L$^3$: Although our calculations have demonstrated that three-body scattering inside the Hill sphere selects for roughly same-sized binaries it is unlikely that this essentially non-linear effect is predictable from two-body considerations. This is because chaos plays a double role: on the one hand it acts as the glue to hold together transient binaries for long periods of time, thereby enhancing their effective capture cross sections. On the other hand, if the intruder itself happens to become entangled in a chaotic zone during its transit through the Hill sphere - an event that is more likely if the primaries have unequal masses - then the binary tends to be either not stabilized or actively destabilized \citep{Astakhov:2005}. Thus, the CAC mechanism discriminates not only between ``big" and ``small" bodies but, actually, also between large and small ``big" bodies. 

\label{sec6}

\begin{figure*}
\begin{center}$
\begin{array}{c}
\includegraphics[scale=0.5]{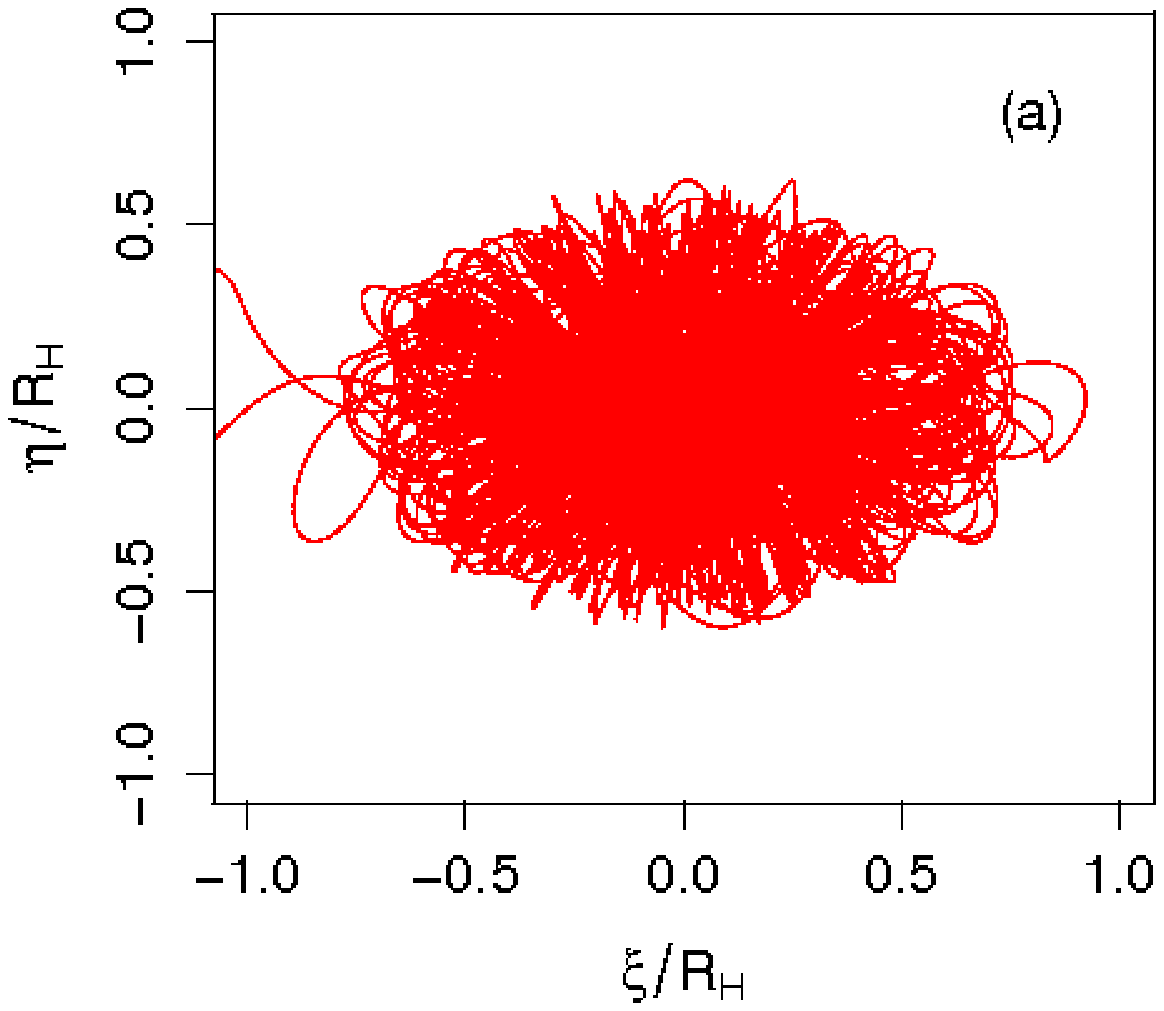}\\
\includegraphics[scale=0.5]{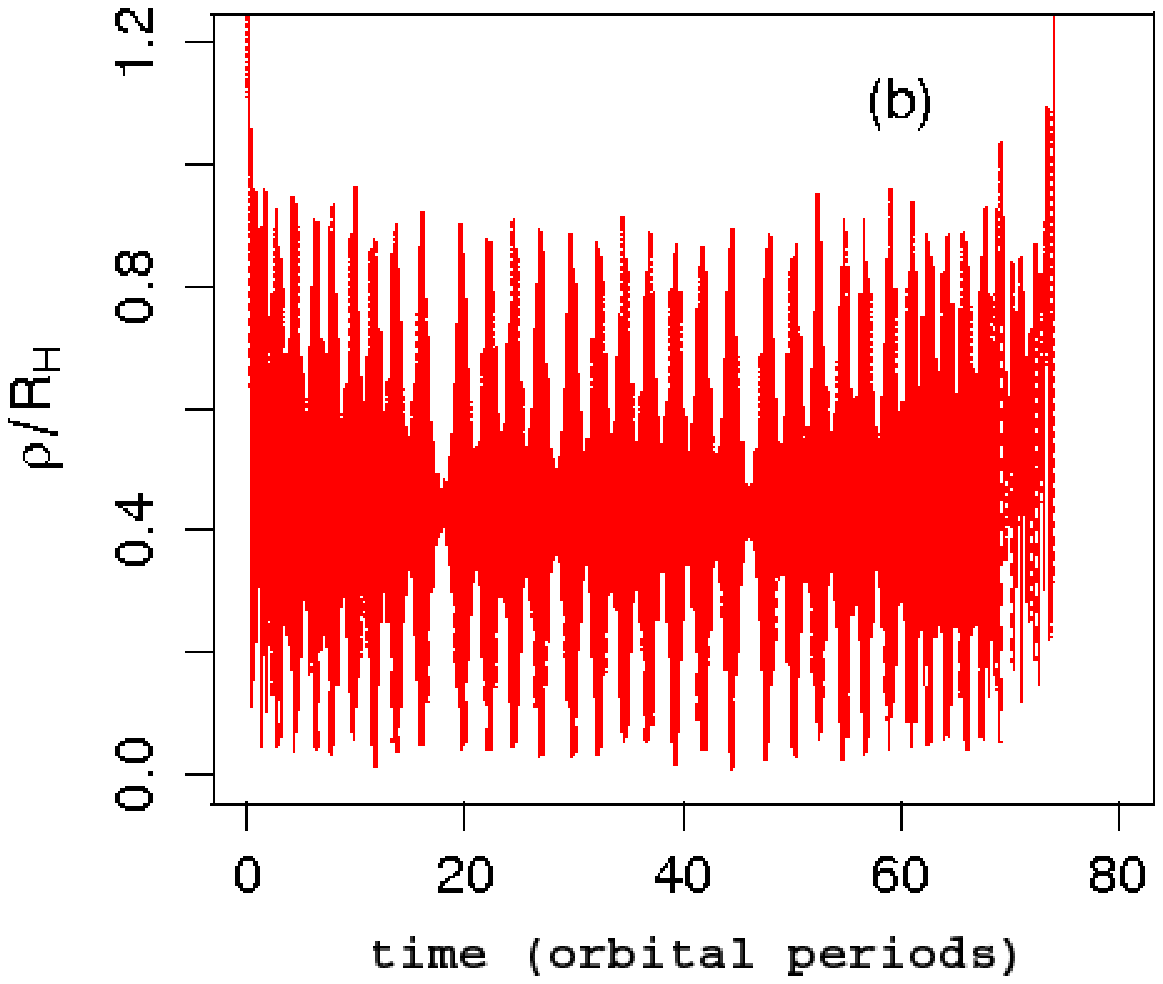}\\
\includegraphics[scale=0.5]{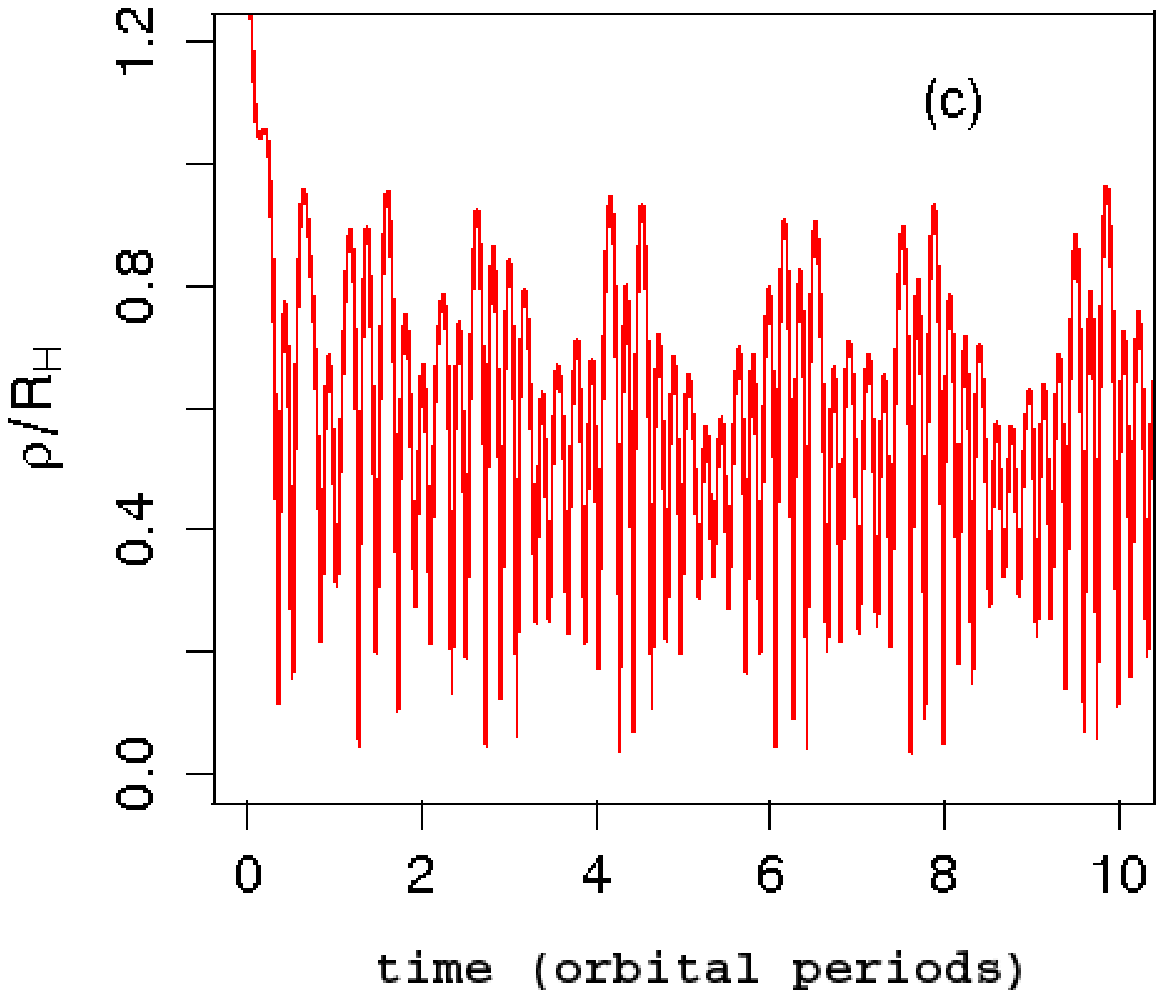}\\

\end{array}$
\end{center}

\caption{\label{traj} An example of a proto-binary trapped in a chaotic layer. Frame (a) shows the $\xi,\eta$ projection of the orbit. Frame (b) shows the radius $\rho$ as a function of time scaled by the orbital period. All distances are scaled by the Hill sphere radius $R_H$. Frame (c) is a blow-up of part of Frame (b). }
\end{figure*}

\section{Binary hardening}
\label{hard}

The question naturally arises as to under what conditions the intruder, on average, hardens - i.e., increases the binding energy - of the binary. We do not address this issue in detail here; for now, we note that various rules of thumb have been developed to understand when a binary hardens or softens in star-binary scattering \citep{Hut:1983,Hut:1983a,Hills:1983,Hills:1983a,Hills:1984,Hills:1990,Heggie:2003,Aarseth:2003,Heggie:1993,Heggie:1996}. For small intruders, \citet{Hills:1990} has argued for a ``fast-slow" rule in which the binding energy of a binary increases (i.e., it hardens) if the velocity of the intruder at infinity is less than the orbital velocity of the binary, and {\it vice-versa.} While large bodies in the disk generally are expected to have smaller velocities than small bodies, in the Hill sphere the binary partners are initially following a chaotic trajectory influenced by solar tides. We find that quasi-binary partner relative velocities are, on average, higher than the asymptotic velocities of incoming intruders. This is due primarily to solar tides (which induce the chaos in the first place). This is illustrated in Fig.\ \ref{traj} which shows the coordinate space projection of a very long lived trajectory. Frames (b) and (c) illustrate that the relative orbital frequency is substantially higher than the heliocentric frequency and, correspondingly, instantaneous (and average) relative velocities can be substantially higher than asymptotic velocities. Therefore, on average, intruder scattering is expected to harden binaries unless resonant triples \citep{Heggie:2003,Aarseth:2003} form in which case the binary tends to be destabilized. This suggests that sequential scattering by relatively small intruders can be an efficient means of orbit reduction, consistent with our previous simulations \citep{Astakhov:2005}.

\section{Limitations of the calculations}
\label{limit}

The calculations reported here have not primarily been an attempt to calculate binary formation rates but, instead, are an investigation into the consequences of the CAC model. Several limitations of the current set of calculations are apparent.
\begin{enumerate}
\item The calculations assume that a quasi-bound binary has already formed prior to the intruder being launched from infinity. Because the mass and orbital element distributions at infinity are poorly characterized then certain attributes of the intruder were selected somewhat arbitrarily, e.g., the initial distance of the intruder from the binary barycenter. While we do not expect the qualitative conclusions of the calculations to be affected, the probability of binary formation will depend on these details. Ideally, initial masses and orbital elements would be selected from a variety of possible mass and velocity distributions and the results compared with each other and with observations. Work to that end is in progress.
\item A hierarchical separation of time scales is assumed in which the binary forms in the three-body Hill approximation and later intruder scattering occurs in the four-body Hill approximation. In principle, quasi-binary formation and intruder scattering could happen essentially simultaneously. However, we expect that such encounters will be relatively infrequent.

\item In these calculations exchange reactions were ignored and the intruder mass was not allowed to exceed the larger of the two primaries. Relaxing these assumptions may lead to modifications of the observed distributions; preliminary investigations indicate that exchange only tends to accentuate the preference for large mass-ratio binaries.

\end{enumerate}

\section{Conclusions}
\label{conclusions}
The model presented in our previous article \citep{Astakhov:2005}, and studied in more detail here, relies on the ability of chaos to produce very long-lived, quasi-bound binaries which can be stabilized by gravitational scattering with an intruder. Roughly equal-sized binary partners are predicted to form which translates into $\Delta_{mag} < 1$. In addition the model suggests that binary formation occurs most readily when the intruder mass is a small percentage of the total binary mass.

Several extensions of the model are possible. For example, it would be interesting to examine the effect of using particular distributions from which to pick initial masses and orbital elements \citep{Greenzweig:1990,Greenzweig:1992,Kenyon:1998,Kenyon:1999,Kenyon:2002,Kenyon:2004}; however, our numerical simulations indicate that a trajectory started randomly in the feeding zone has $\sim 0.8 \%$ chance of actually entering the Hill sphere - and a smaller probability of subsequently becoming caught up in a chaotic layer. This makes numerical simulations very inefficient in that most integrations result in no possibility of capture. That is, it is difficult, in practice, to select initial conditions directly in the asymptotic region because most do not lead to interpenetration of the Hill sphere. 

To address these issues and to increase the efficiency of the scattering calculations we are in the process of developing a method which uses random forests of decision trees
\citep{Breiman:1984} as a Boolean classifier to discriminate between instances - sets of initial conditions at infinity - which lead to primary encounters inside the Hill sphere (positives) and those which do not (negatives).

To understand the capture dynamics further and to predict formation rates it may be fruitful to employ phase-space transport theory \citep{Jaffe:2000,Jaffe:2000a,Jaffe:2002,Uzer:2002} to study both the initial temporary trapping of the primaries and their subsequent stabilization or destabilization through intruder scattering. Essentially the procedure is as follows: (i) identify the critical saddle-points (in the Hill problem these are the Lagrange points $L_1$ and $L_2$), (ii) make a Birkhoff normal form expansion at the saddle points, (iii) construct the normally invariant hyperbolic manifolds and the transition state(s) from the normal form \citep{Wiggins:1994}. The transition state is a dividing surface in phase space that acts a surface of no return. This method has already been applied to the three-body Hill problem, although only over a quite limited energy range \citep{Waalkens:2005,Waalkens:2005a,Waalkens:2005b}.

New observations will provide rigorous tests of the CAC scenario. Observationally, few of the already known TNBs are fully characterized. In particular the true inclination of the mutual orbital plane generally remains unknown. Because of the small parallax, ambiguities in the orbit inclination of TNBs will persist until observations are performed at well separated epochs \citep{Hestroffer:2005}. Knowledge of true inclinations is especially important for predicting stellar occultations and mutual phenomena from which binary densities can be derived \citep{Brown:1997,Pravec:1998,Cooray:2003}. Orbital inclinations are of interest here for two reasons: (i) the CAC formation model predicts asymmetric distributions of prograde {\it vs.\  }retrograde orbits with respect to the invariant plane; (ii) the numerical results presented here and actual observations of inclination distributions are consistent with TNB production during the oligarchic growth phase which comes directly after a shorter, runaway accretion regime in the model of \citet{Rafikov:2003}.

In summary, trans-Neptunian binaries present multiple conundrums which include their unexpectedly high proportion in the total population of TNOs; their large, mid-range eccentricity, orbits; their mutual orbital inclination distributions; and their apparent propensity to have roughly equal-sized partners. Resolving these questions will undoubtedly shed light on the formation of the trans-Neptunian part of the Solar System, with implications for planetary formation theories in general \citep{Youdin:2003,Goldreich:2004,Basri:2006}. The results presented here suggest that chaos might have played an important - and constructive - role in the formation of these remote and puzzling objects.

\section*{acknowledgments}
We thank Keith Noll for providing us with an advance copy of a chapter to appear in the book {\sl The Kuiper Belt} \citep{Noll:2007}. We are grateful to Kevin Hestir, Daniel Hestroffer  and Susan Kern for their suggestions and comments.  We also thank the referee for a number of helpful and insightful comments. Extensive use has been made of the R statistical package for analysis and graphics \citep{rref,hmisc,brewer,biobase}. This work was supported, in part, by a grant from the National Science Foundation (USA).

\appendix

\section{Kernel Smoothing}

Scatterplots are a useful way of visualizing data, e.g., the data in Fig.\ 2 could be represented simply as a scatterplot of $\log_{10}(T_{\text{Hill}})$ against velocity. However, when there exists a very high density of points in certain localized regions it may become difficult to distinguish local density gradients  \citep{Gentleman:2006,Carr:1987,gene}. In kernel smoothing the data is divided into two-dimensional bins and the bin coloured according to the local density using a false-colour representation. A kernel density estimator is used to estimate the local density of points and to remove the dependence on the end points of the bins.  The estimator centers a kernel function at each data point, e.g., a Gaussian. The kernel estimator essentially smears the contribution of each data point over a local region which depends on the bandwidth and which, in turn, determines the relative contributions of adjacent points. Visual comparison of the raw scatterplot at high resolution and the kernel-smoothed plots shows that kernel smoothing enhances nuances in the density variation that are hard to see in the scatterplot at the same resolution. Standard routines are available in the R statistical package for analysis and graphics \citep{rref,hmisc,brewer,biobase} to make these plots. Because this is primarily a visualization procedure it is difficult to assign a quantitative colour scale; just as in a conventional scatterplot the density of points is best judged by eye.

A simple analogue might be using a moving average to smooth a time series in one-dimension and then making a histogram of the moving average rather than of the raw data.

\bibliographystyle{mn2e}
\bibliography{kbb2}

\end{document}